\documentclass[preprint, twocolumn]{aastex631}
\graphicspath{{./}{Figures/}}

\usepackage{verbatim}

\begin{document}

\title{Structural Parameters of the Thin Disk Population from Evolved Stars in Solar Neighborhood}

\correspondingauthor{Sel\c{c}uk Bilir}
\email{sbilir@istanbul.edu.tr}

\author[0009-0002-2272-8898]{Sedanur İyİsan}
\affiliation{Istanbul University, Institute of Graduate Studies in Science, Programme of Astronomy and Space Sciences, 34116, Istanbul, Turkey}

\author[0000-0003-3510-1509]{Sel\c{c}uk Bİlİr}
\affiliation{Istanbul University, Faculty of Science, Department of Astronomy and Space Sciences, 34119, Istanbul, Turkey}

\author[0000-0003-0864-1921]{Özgecan Önal Taş}
\affiliation{Istanbul University, Faculty of Science, Department of Astronomy and Space Sciences, 34119, Istanbul, Turkey}

\author[0000-0002-0435-4493]{Olcay Plevne}
\affiliation{Istanbul University, Faculty of Science, Department of Astronomy and Space Sciences, 34119, Istanbul, Turkey}



\begin{abstract}
This study investigates the structural parameters of the thin-disk population by analyzing the spatial distribution of evolved stars in the solar neighbourhood. From the $\it Gaia$ Data Release 3 database, about 39.1 million stars within 1 kpc and with relative parallax errors $\sigma_{\varpi}/\varpi\leq 0.10$ were selected. The photometric data was corrected for extinction using a Galactic dust map. The sample was refined by considering the color-magnitude region $M_{\rm G}\times (G_{\rm BP}-G_{\rm RP})_0$ associated with evolved stars, applying a stricter parallax error limit of $\sigma_{\varpi}/\varpi\leq 0.02$, and yielding 671,600 stars. The star sample was divided into 36 regions based on their Galactic coordinates, with evolved stars in the absolute magnitude range of $-1< M_{\rm G}~{\rm (mag)}\leq 4$ further split into five one-unit magnitude intervals. This led to 180 subgroups whose space density profiles were modelled using a single-component Galaxy model. The analysis shows that the space densities are in agreement with the literature and that the scale heights vary with $200<H~{\rm (pc)}<600$ interval to their absolute magnitudes. Red clump stars in the solar neighbourhood were also estimated to have a scale height of $295\pm10$ pc. These findings indicate that evolved stars with bright absolute magnitudes originate from the evolution of the early spectral-type stars with short scale height, while fainter ones come from the evolution of the intermediate spectral-type stars with large scale height, suggesting variations in scale height reflect the contribution of Galactic evolution processes.
\end{abstract}



\section{Introduction} \label{sec:intro}

The study of Galactic structure has been essential in advancing our understanding of the universe. Detailed analysis of the Milky Way allows us to address key astrophysical questions, as we can observe it more closely than any other galaxy. Techniques like star counting, alongside methods incorporating stellar chemistry, age, and kinematics, were particularly useful for determining the Galaxy's structure \citep{Peiris2000}. In particular, the star counting method helps constrain the Galaxy's components, assuming a density distribution similar to galaxies of the same Hubble type. According to the standard model, the Milky Way is a Hubble-type Sbc galaxy with an exponential disk and a spheroidal component \citep{Binney1998}.

Since the 1980s, substantial progress has been made in Galactic structure research, particularly through combining observational data with theoretical models. \citet{Bahcall1980} developed the first consistent model using star count data, revealing that the Milky Way consists of two Galactic components as the thin disk with a scale height of 325 pc and the halo. However, \citet{Gilmore-Reid1983} proposed a three-component model, adding the thick disk to account for discrepancies in the data. According to \citet{Gilmore-Reid1983} the thick disk has a scale height of 1450 pc for $1<z~{\rm (kpc)}<5$ distance range and suggested the thick disk contributes 2\% of the thin disk's density in the solar neighbourhood.

Subsequent studies have further refined the Galactic model parameters of the Milky Way. \cite{Ojha1996} estimated that the vertical scale height of stars fainter than $M_{\rm V}=3.5$ mag in the thin disk was $h_{\rm z}=260\pm 50$ pc, while in the thick disk scale height was $h_{\rm z}=760\pm 50$ pc, representing 7.4\% of the local density of the thin disk. These findings have been instrumental in shaping our current understanding of the multi-component structure of the Galaxy. \citet{Buser1998, Buser1999} studied star counts using \textit{RGU} photometric data of star fields in different directions of the Galaxy within the scope of the Basel Palomar-Schmidt programme. As a result, they estimated the local density of the thick disk to be $5.9\pm 3\%$ and the scale length and scale height of the thick disk to be $3.0\pm 1.5$ kpc and $0.91\pm 0.3$ kpc, respectively. Although the presence of the thick disk was well established, the Galaxy model parameters of the three Galactic components, especially the thick disk population, could not be given clearly \citep{Chen2001, Siegel2002}.

With the development of technology, the common use of CCDs in astronomical observations has enabled fainter stars to be analysed \citep[e.g.,][]{Hall1996, Karaali2003}. Thanks to the systematic sky surveys that started in the 2000s, a new era has entered the study of the three-dimensional structure of the Galaxy. \citet{Karaali2004}, who analyzed the SA114 star field with the 2.5 m INT telescope using $u'_{\rm RGO}g'r'i'z'_{\rm RGO}$ filters\footnote{RGO denotes Royal Greenwich Observatory}, showed that the scale height of the thin-disk population increased from 265 pc to 495 pc as the absolute magnitude changed from bright to faint. Additionally, the density of the thick disk component in the solar neighbourhood decreased from 9.5\% to 5.2\% with the same change in absolute magnitude. However, they also found that the halo star density in the solar vicinity was in the range of 0.02-0.05\% and the halo axial ratio was $c/a=0.7$. \citet{Bilir2006a}, who studied the ELAIS star field with photometric data from the INT 2.5m telescope, obtained the scale height for the thin disk as $H=269\pm 8$ pc, the star density of the thick disk as 6.46\% and the scale height as $H=760\pm 60$ pc, the star density of the halo in the solar neighbourhood as 0.08\% and the axis ratio as $c/a=0.55\pm 0.20$. Photometric studies of relatively small star fields in different directions of the Galaxy have shown that unique parameters describing the Galactic population have not yet been obtained \citep[see also,][]{Du2003, Bilir2006b, Bilir2006c, Ak2007a, Ak2007b, Wang2018, Yu2021, Chrobakova2020}.  

\citet{Juric2008} provided Galaxy model parameters for three Galactic populations from the analysis of Sloan Digital Sky Survey \citep[SDSS,][]{York2000} photometric data of about 48 million stars in a region of about 6,500 square degrees of sky. As a result of the analyses of M dwarf stars in the solar neighbourhood, the scale length and scale height of the thin disk were determined as $h_1 = 2600$ pc and $H_1=300$ pc, respectively. Also, the local stellar density, scale length and scale height of the thick disk were found as 12\%, $h_2 = 3600$ pc and $H_2=900$ pc, respectively, while the local stellar density and axis ratio of the halo were estimated as 0.5\% and $c/a=0.64$, respectively. Similar studies have also been conducted with different objects at high Galactic latitudes. \citet{Cabrera-Lavers2007} analysed the Two Micron Sky Survey \citep[2MASS,][]{Skrutskie2006} photometry data of red clump stars ($0^{\rm o}<l\leq 360^{\rm o}$, $60^{\rm o}<b\leq70^{\rm o}$), which were good distance indicators, while \citet{Bilir2008a} studied the SDSS photometric data of main-sequence stars with different absolute magnitude intervals in the almost sky region ($0^{\rm o}<l\leq 360^{\rm o}$, $60^{\rm o}<b\leq65^{\rm o}$). Although both studies were carried out in nearly the same region of the sky, it was shown that the Galaxy model parameters for Galactic populations were defined over a wide range of parameters. 

The wide range of Galaxy model parameters may be attributed to many factors, with the main reasons organized as follows: i) Galaxy model parameters depend on the Galactic latitude/longitude of the sources \citep{Bilir2006a, Bilir2006b, Bilir2006c, Cabrera-Lavers2007, Ak2007a, Yaz2010, Yaz2015}. ii) Galaxy model parameters depend on the luminosity class of stars with wide absolute magnitude ranges \citep{Karaali2004, Bilir2006a, Bilir2006b, Bilir2006c, Cabrera-Lavers2007, Bilir2008a}. iii) Galaxy model parameters have different values in different space volumes depending on the completeness limits \citep{Karaali2007}. iv) In the star counting method, the distances of the stars were determined by different photometric parallax calibrations \citep{Karaali2005, Bilir2005, Bilir2008b, Bilir2009}. v) Depending on the stellar sample used in Galaxy model parameter determination studies, the morphology of the density functions used to separate the thin disk, thick disk, and halo components of the Galaxy varies. In the literature, studies have demonstrated that stars with bright absolute magnitudes are well represented by an exponential law, whereas samples of stars with faint absolute magnitudes are more accurately described by the $\sec$ or sech$^2$ law \citep{Bilir2006a, Bilir2006c, Karaali2009, Yaz2010, Yaz2015}. The Galaxy model parameters cannot be expressed uniquely for the reasons listed above.

The {\it Gaia} mission aims to create a complex three-dimensional map of the Milky Way through precise measurements of the positions, distances and motions of more than a billion stars \citep{Gaia2016a}. For this purpose, the {\it Gaia} satellite was launched by the European Space Agency (ESA) in 2013. The {\it Gaia} catalogue, with three main data releases so far, contains astrometric data such as positions, trigonometric parallax, and proper motion components of 1.8 billion stars; photometric data including magnitude in the $G$, $G_{\rm BP}$, and $G_{\rm RP}$ bands defined at optical wavelengths; and spectroscopic data including atmospheric model parameters and radial velocity measurements from low-resolution spectra \citep{GaiaDR3}. With these features, the {\it Gaia} database allows us to study the structure of the Milky Way in detail. 

Using precise astrometric data from the {\it Gaia} EDR3 \citep{GaiaEDR3} catalogue, \citet{Gaia-Smart2021} obtained space densities and luminosity functions by classifying 331\,312 stars within 100 pc into dwarf, giant and white dwarf. As a result of this study, the luminosity functions of main-sequence stars have been obtained in the absolute magnitude $-1\leq M_{\rm G}~{\rm (mag)}\leq 20$ interval and the luminosity functions of red giant stars have been obtained for the range of $-1<M_{\rm G}~{\rm (mag)}\leq 4$. Especially in the range $0<M_{\rm G}~{\rm (mag)}\leq 1$, where red giants stars were found, the luminosity function as $\Theta=1.9\pm0.1\times10^{-4}$ stars pc$^{-3}$ mag$^{-1}$ was found to increase significantly. Since the results were obtained for a very limited volume of space, other Galaxy model parameters such as the scale length and scale height could not calculated. 

The study of evolved stars in the solar neighbourhood with precise data provided by the {\it Gaia} satellite was of great importance in estimating the Galaxy model parameters of the thin-disk population. In this study, we used photometric and astrometric data from the {\it Gaia} satellite to identify evolved stars within 1 kpc of the Sun using a color-magnitude diagram. The Galaxy model parameters were estimated by dividing the selected evolved stars into different Galactic latitude and longitude intervals and different absolute magnitude ranges.

\section{The Density-Law of the Thin Disk}
Disk structures in Galaxy models, especially in the studies of Galactic disks, are usually parameterized in cylindrical coordinates through radial and vertical exponential functions. This parametrization allows for a more precise representation of the density distribution along both the radial and vertical axes related to the Galactic plane. The double exponential density law for the thin disk of the Milky Way is represented as follows.

\begin{eqnarray}
D(x,z)=n\times \exp \left(-\frac{|z+z_0|}{H} \right)\times \exp \left(-\frac{(x-R_{0})}{h}\right)
\label{eq: Eq1}
\end{eqnarray}
where $x$ is the planar distance from the Galactic centre, $z$ is the distance of stars from the Galactic plane, $z_0$ distance of the Sun from the Galactic plane \citep[15 pc,][]{Cohen1995,Hammersley1995}, $R_{0}$ is the solar distance to the Galactic centre \citep[8 kpc,][]{Majewski1993}, $n$ is the normalized local density, $h$ and $H$ are the scale length and scale height of the thin-disk component, respectively. The following equation calculates the planar distance from the Galactic centre. 
\begin{eqnarray}
x = [R_{0}^{2}+(z/\tan b)^2-2R_{0}(z/\tan b)\cos l]^{1/2}
\label{eq: Eq2}
\end{eqnarray}
here $l$ and $b$ are the Galactic longitude and latitude of the star field under study. Since this study analyzes evolved stars in the solar neighbourhood ($d\leq 1$ kpc), the sample does not reach far enough to determine the scale length of the thin disk. Therefore, Equation ~\ref{eq: Eq1} is simplified to determine only the space densities and scale heights of thin disk stars.
 
\begin{eqnarray}
D(z)\cong n\times \exp \left(-\frac{|z+z_0|}{H} \right)
\label{eq: Eq3}
\end{eqnarray}

\section{Data and Analyses}
The {\it Gaia} Third Data Release \citep[{\it Gaia} DR3,][]{GaiaDR3} contains astrometric, photometric, and spectroscopic data from 34 months of observations of 1.8 billion sources. The precision of the trigonometric parallaxes in the database varies depending on the $G$-apparent magnitudes. The trigonometric parallax errors were $\sigma_{\varpi}=0.03$ mas for $G<15$ mag, $\sigma_{\varpi}=0.07$ mas for $G=17$ mag and $\sigma_{\varpi}=0.5$ mas for $G=20$ mag were given in the {\it Gaia} DR3 database. Similarly, the uncertainties of the proper motion measurements were 0.03 mas yr$^{-1}$ for $G<15$ mag, 0.07 mas yr$^{-1}$ for $G=17$ mag and 1.40 mas yr$^{-1}$ for $G=21$ mag. 

This study uses the photometric and astrometric data of the evolved stars in {\it Gaia} DR3 database \citep{GaiaDR3} to determine the Galaxy model parameters of the thin-disk population. The photometric and astrometric data in the \texttt{gaiaedr3.gaia\_source} table\footnote{\protect\url{https://gaia.aip.de/metadata/gaiadr3/gaia_source/}} were used to select the stars in the 1 kpc heliocentric space volume. To determine the distances of the stars more accurately and sensitively, it was considered appropriate to consider single stars with relative parallax errors $\sigma_{\varpi}/\varpi\leq 0.10$ in the {\it Gaia} DR3 database. In addition, the value of 0.017 mas was taken into account for the global parallax offset in the trigonometric parallax data of the selected stars \citep{Lindegren2021}. For the selection of single stars in the stellar sample, the Renormalised Unit Weight Error (RUWE) parameters given in the {\it Gaia} DR3 database for each source were chosen such as RUWE $\leq$ 1.4 \citep[e.g.][]{Castro-Ginard2024}. Considering the main points mentioned above, the following code was written in the SQL section of the {\it Gaia} DR3 database and the data providing these conditions were listed. 

\begin{verbatim}
print("SELECT *
FROM gaiadr3.gaia_source as gaia
WHERE gaia.parallax_over_error >=10 AND
      (gaia.parallax+0.017) >=1 AND 
      RUWE <= 1.4")
\end{verbatim}
This SQL query resulted in 39,099,903 stars within the 1 kpc distance that have relative parallax errors of less than 0.10. The distances of the stars were calculated with the conventional inverse parallax relation, $d{\rm (pc)}=1000/\varpi$ (mas), using {\it Gaia} DR3 data. Photometric {\it Gaia} bands were dereddened using the below methodology. 

\subsection{Photometric Color Excess Determination}
In this study, the dust map of \cite{Schlafly2011} was used to correct the photometric data of selected stars from the {\it Gaia} DR3 catalogue for the extinction effect of the interstellar medium. The $V$-band extinction ($A_{\infty}(V)$) values in the line of the stars ($l$, $b$) up to the Galactic boundary were determined with the help of the python library \texttt{mwdust} \citep{Bovy2016a}. Since the stars in this study were not located at the Galactic boundary, the $V$-band extinction value ($A_{\rm d}(V)$) determined from the dust map of \cite{Schlafly2011} needs to be recalculated for the distance between the Sun and stars. The relation of \cite{Bahcall1980} was used to determine the reduced $V$-band extinction.

\begin{equation}
A_{\rm d}(V)=A_{\infty}(V)\times \left[1-\exp\left(\frac{-|d\times\sin b|}{H}\right)\right]
\label{equ: Eq4}
\end{equation} 
where $b$ is the Galactic latitude of the star, $d$ is the distance of the star which is calculated from the corrected trigonometric parallax measurements of the star in the {\it Gaia} DR3 catalogue with the relation $d{\rm (pc)}=1000/\varpi$ (mas), $H$ is the scale height of the dust \citep[$H=125^{+17}_{-7}$ pc,][]{Marshall2006}, $A_{\infty}(V)$ is the $V$-band extinction measured from the star line to the Galaxy boundary, and $A_{\rm d}(V)$ is the extinction value for the distance between the Sun and star. 

\begin{figure}
\centering
\includegraphics[width=\columnwidth]{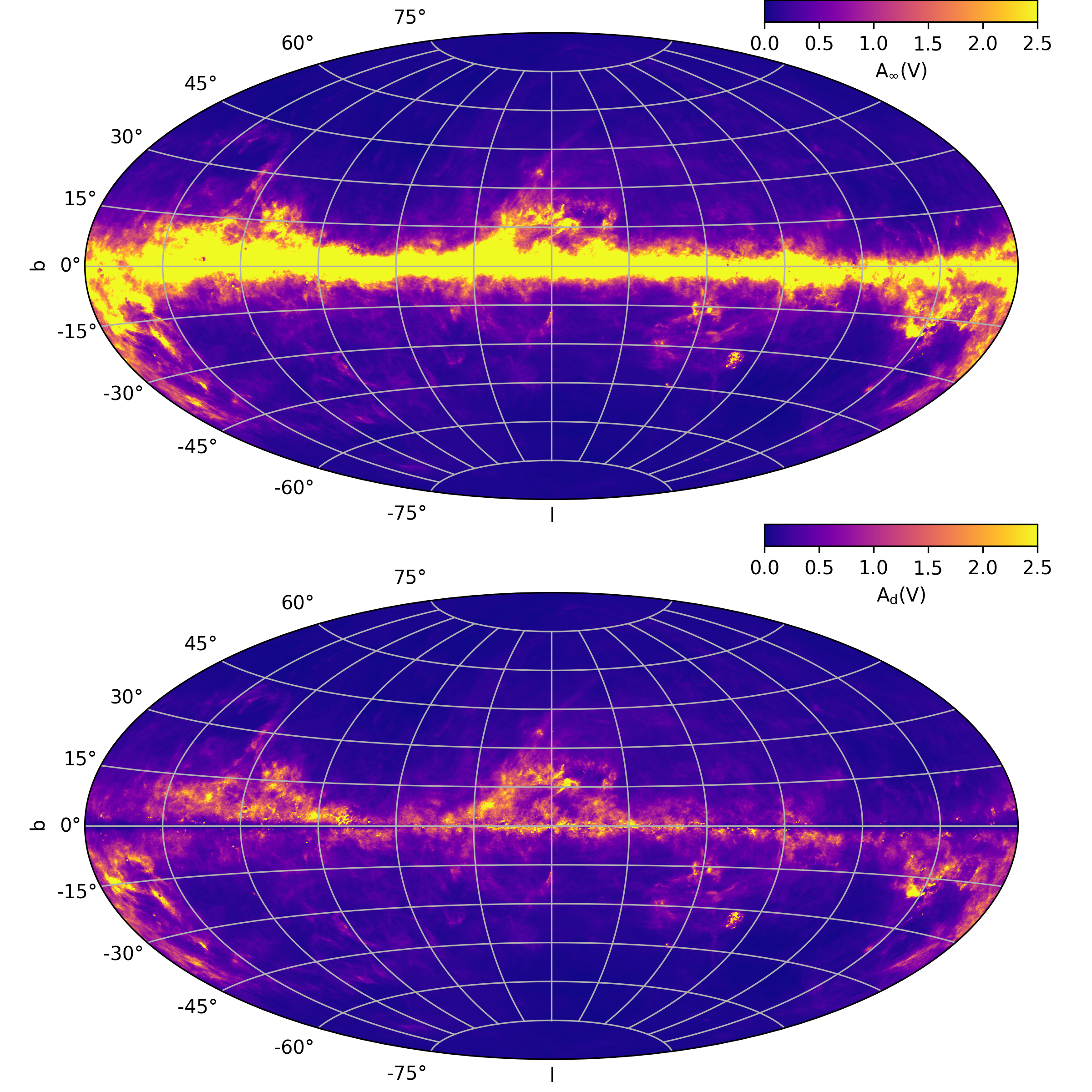}
\caption{$V$-band extinction $A_{\infty}(V)$ values from the dust maps of \citet{Schlafly2011} (upper panel) and the distance between Sun and stars reduced $A_{\rm d}(V)$ values (lower panel).} 
\label{fig:Fig01}
\end{figure}

The selective absorption coefficients of \cite{Cardelli1989} were used to correct the photometric bands in the {\it Gaia} DR3 catalogue for the effect of interstellar extinction. In this study, the $R_V=3.1$ curve of \cite{Cardelli1989} was used to determine the extinction coefficients of these filters. The effective wavelengths of {\it Gaia} passbands for $G$, $G_{\rm BP}$ and $G_{\rm RP}$ are 6390.21 \AA, 5182.58\AA~and 7825.05\AA, respectively, the corresponding $A_{\lambda}/A_{\rm V}$ values are 0.83627, 1.08337 and 0.63439, respectively \citep[see also,][]{Canbay2023}. Accordingly, the relations that should be used for de-reddening three {\it Gaia} passbands are as follows:

\begin{eqnarray}
G_{\rm 0} = G-A_{\rm G} = G - 0.83627\times\ A_{\rm d}(V)\\ \nonumber
(G_{\rm BP})_{0} = G_{\rm BP}-A_{\rm G_{\rm BP}} = G_{\rm BP} - 1.08337\times\ A_{\rm d}(V) \\ \nonumber
(G_{\rm RP})_{0} = G_{\rm RP}-A_{\rm G_{\rm RP}} = G_{\rm RP} - 0.63439\times\ A_{\rm d}(V) \\  \nonumber
\label{equ: Eq5}
\end{eqnarray}
The absolute magnitudes $M_{\rm G}$ of stars were calculated using the distance modulus formula $G-M_{\rm G} = 5\times \log(1000/\varpi)-5+A_{\rm G}$, where $\varpi$ is the trigonometric parallax with global zero point corrected. The $V$-band extinctions from the \cite{Schlafly2011} dust maps for the selected stars were shown in the upper panel of Figure~\ref{fig:Fig01}, and calculated for the distance between the Sun and the star in $V$-band absorption were also represented in the lower panel of Figure~\ref{fig:Fig01}, as well.

\subsection{Selection of Evolved Stars}
\begin{figure}
\centering
\includegraphics[width=\columnwidth]{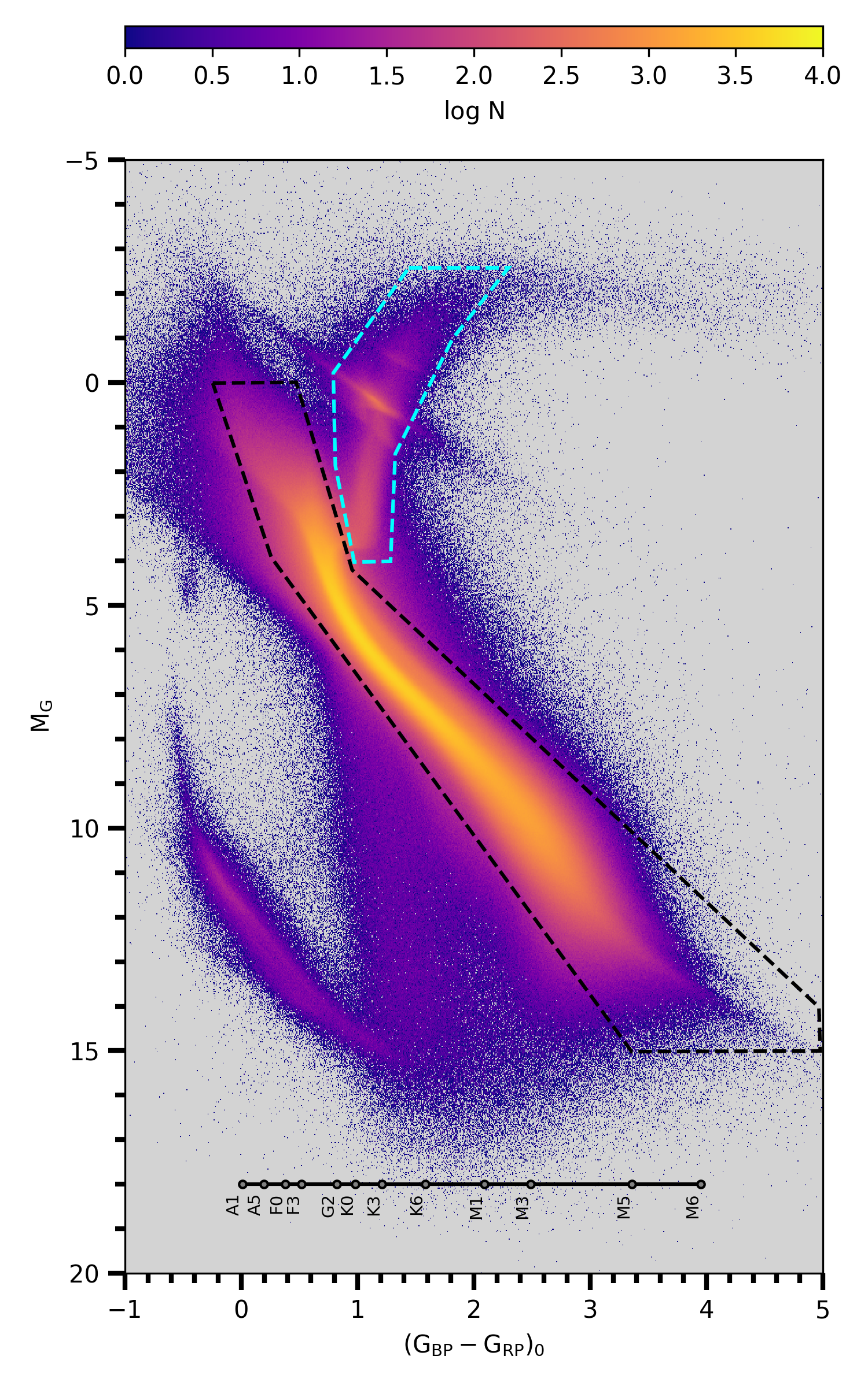}
\caption{$M_{\rm G}\times(G_{\rm BP}-G_{\rm RP})_0$ CMD of a sample of 39,099,903 stars within the solar neighbourhood. Main sequence and evolved stars occupy the black dashed and blue dashed regions, respectively. Another star clump at the lower left part of the diagram was the white dwarf region. Spectral types concerning the de-reddened color index were shown at the bottom part of the diagram.} 
\label{fig:Fig02}
\end{figure}

The color-magnitude diagram (CMD) of 39,099,903 stars of different luminosity classes in a heliocentric 1 kpc volume of space was shown in Figure \ref{fig:Fig02}. To better represent the star densities on the $M_{\rm G}\times (G_{\rm BP}-G_{\rm RP})_0$ CMD, the stars were colored according to their number densities. The main-sequence, evolved and white dwarf stars in the sample have different positions on the CMD and can be easily distinguished by the eye. In addition, the spectral types of the stars given by the web page of Mamajek\footnote{\protect\url{http://www.pas.rochester.edu/~emamajek/EEM_dwarf_UBVIJHK_colors_Teff.txt}} for {\it Gaia} photometry were shown at the lower part of Figure \ref{fig:Fig02}. Considering these properties of the star groups, main-sequence stars and evolved stars were separated by green and red dashed lines as shown in Figure \ref{fig:Fig02}, respectively. The color indices and absolute magnitudes of main-sequence stars span $-1<(G_{\rm BP}-G_{\rm RP})_0~{\rm (mag)} <4$ and $-2< M_{\rm G}~{\rm (mag)}<15$, while evolved stars cover $0.8< (G_{\rm BP}-G_{\rm RP})_0~{\rm (mag)}<2.2$ and $-3< M_{\rm G}~{\rm (mag)}<4$ intervals. The number of evolved stars in the area bounded by the blue dashed lines on the figure showing the evolved star region was identified as 776,246. This evolved star region is also composed of sub-evolved classes such as red giant branch (RGB; $-3<M_{\rm G}~{\rm (mag)}<4$), red clump (RC; $M_{\rm G}=0.5$, $G_{\rm BP}-G_{\rm RP})_0=1.2$~mag), secondary red clump ($M_{\rm G}=0.6$, $G_{\rm BP}-G_{\rm RP})_0=1.1$~mag) and asymptotic giant branch (AGB) bump stars \citep[see also,][]{GaiaDR2-HR}. 

\begin{figure}
    \centering
    \includegraphics[width=\columnwidth]{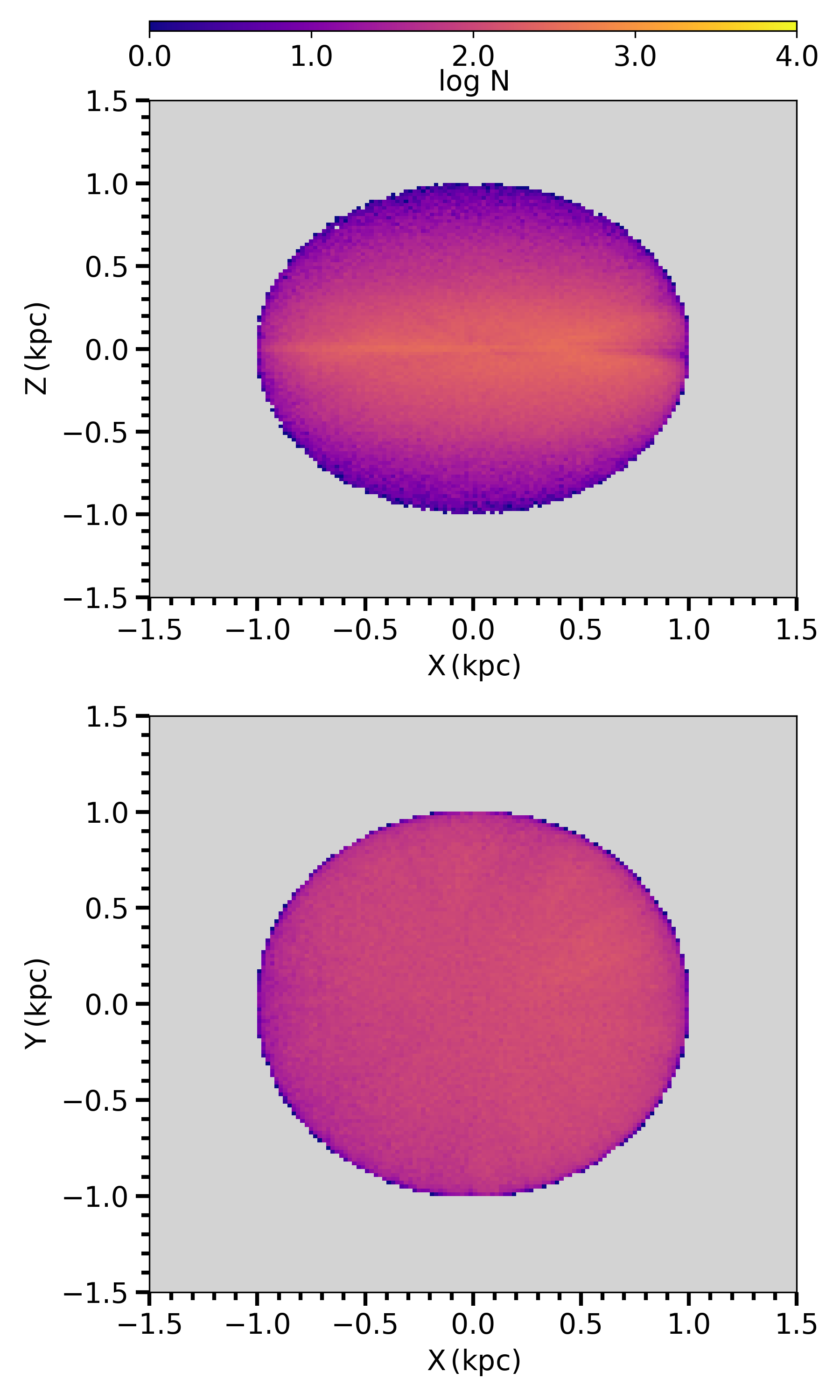}
    \caption{Spatial distributions of 671,600 evolved stars whose astrometric data were precisely selected from the {\it Gaia} DR3 catalogue: $Z \times X$ (upper panel) and  $Y \times X$ (below panel). The color scale indicates the density of the star count.}
    \label{fig:Fig03}
\end{figure}

\subsection{Spatial Distributions and Completeness Limits}

Determining distance and completeness limits for the objects used in calculating the Galaxy model parameters is critical for ensuring accuracy and precision for the unique parameters. In this study, the distances of selected evolved stars from the Sun were determined by applying the trigonometric parallax data provided in the {\it Gaia} DR3 catalogue to the standard distance-parallax relation, $d{\rm (pc)}=1000/\varpi$ (mas), where $d$ is the distance and $\varpi$ is the corrected trigonometric parallax. While the relative parallax error ($\sigma_{\varpi} / \varpi$) for the stars in the selected sample was set at 0.10, it was found that the most reliable distances were constrained within a relative parallax error of 0.02. Within this limit, approximately 86\% of the selected evolved stars were included. By applying this constraint to the star sample, the total number of stars considered was 671,600. To investigate the spatial distribution of the sample, we calculated the heliocentric rectangular Galactic coordinates ($X$ towards the Galactic Center, $Y$ Galactic rotation, $Z$ North Galactic Pole). Figure \ref{fig:Fig03} displays the projected positions on the Galactic plane ($X, Y$) and the Galactic plane perpendicular to it ($X, Z$). In particular, the stellar density in the $Y\times X$ plane shows a homogeneous distribution of stars around the Sun, while in the $Z\times X$ plane there was a slight deviation from the homogeneous distribution. 

To determine the completeness distances of the stars in the sample, the apparent and absolute magnitudes of the stars must be known. The apparent and absolute magnitudes of the selected evolved stars were in the $3<G~{\rm (mag)}\leq 22$ and $-1<M_{\rm G}~{\rm (mag)}\leq 4$ intervals, respectively. In the study, the completeness limits of the stars in the volume of space in which they are located were determined by dividing the stars into intervals of unit absolute magnitude. The following relations were used to estimate the completeness distance limits.

\begin{eqnarray}
 \label{eq: Eq6}
 d_{\rm min}=10^{[(G_{\rm 1}-M_{1}+5-A_{\rm G})/5]} \\ 
 \nonumber
 d_{\rm max}=10^{[(G_{\rm 2}-M_{2}+5-A_{\rm G})/5]}
\end{eqnarray}
Here, $G_1$ and $G_2$ represent the brightest and faintest de-reddened apparent magnitudes within the specified range of absolute magnitudes (e.g. $M_1<M_{\rm G}\leq M_2$), while $d_{\rm min}$ and $d_{\rm max}$ correspond to the minimum and maximum limiting distances, respectively. 

\begin{figure*}
\centering
\includegraphics[width=\textwidth]{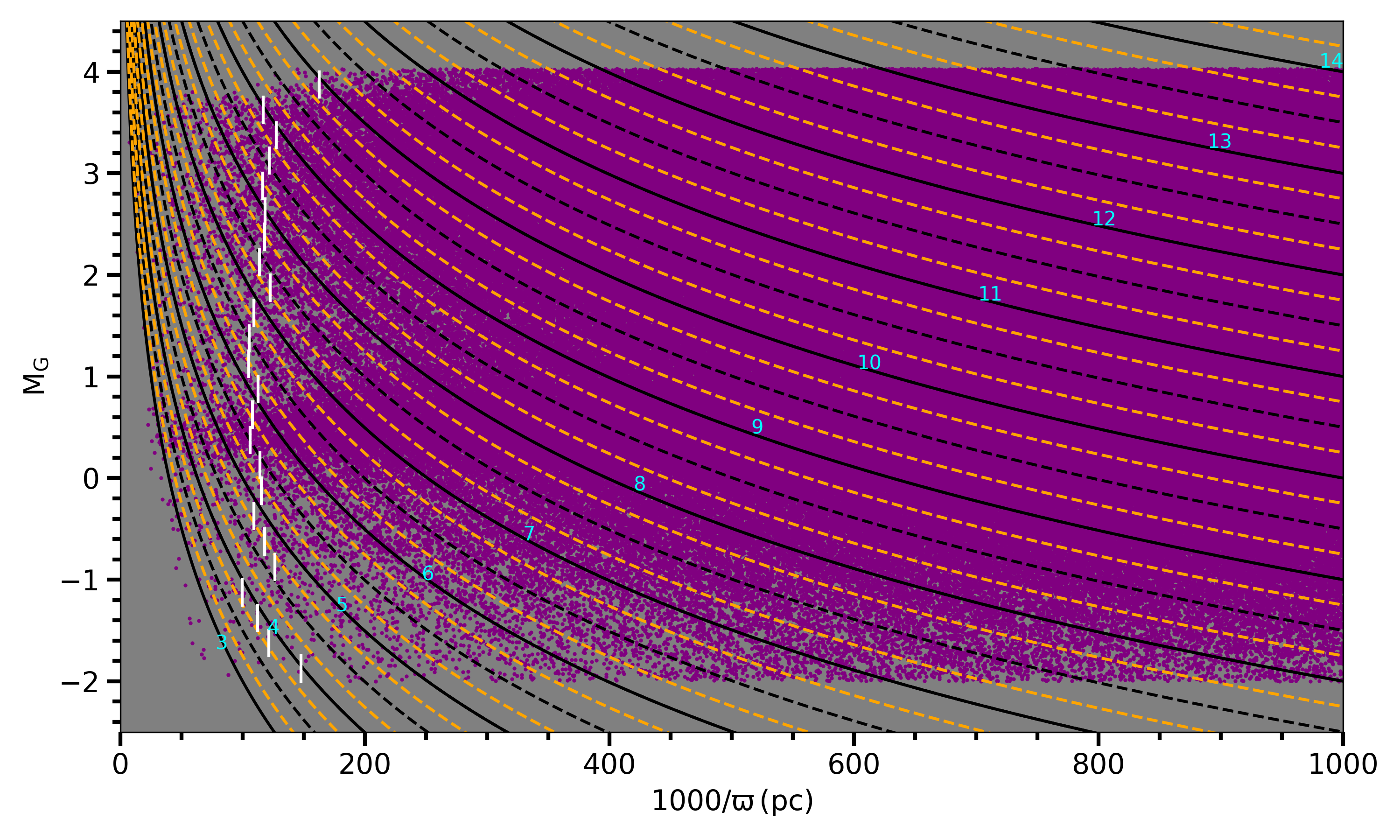}
\caption{$M_{\rm G} \times d$ diagram of evolved stars with $\sigma_{\varpi} / \varpi \leq$0.02. White vertical lines were bright $G$-apparent magnitude limits for different $M_{\rm G}$ absolute magnitude intervals, and solid and dash curves also represent $G$-apparent magnitudes.} 
\label{fig:Fig05}
\end{figure*}

To evaluate the completeness limits of the star sample, we constructed a diagram representing the absolute magnitude in the {\it Gaia} $G$-band in Figure~\ref{fig:Fig05}, with completeness limits marked at intervals of 0.25 magnitudes. The vertical white lines in Figure~\ref{fig:Fig05} signify the distance at which completeness begins for each absolute magnitude interval. We determined these starting points by identifying the initial 0.5\% slice of the $G$-band apparent magnitude distribution within each absolute magnitude bin. Additionally, the black solid lines labelled with turquoise numbers represent the $G$-apparent magnitudes, while the black dashed lines denote increments of 0.5 magnitudes between two consecutive $G$-apparent magnitudes. The orange dashed lines further subdivide these increments, indicating the 0.25 and 0.75 magnitude levels. For subsequent calculations of space density, data points lying below this completeness threshold were excluded from the sample. To determine the faint limiting magnitude of the evolved stars in the sample, the $G$-apparent magnitudes of stars with different absolute magnitudes at a distance of $d=1$ kpc were used. Accordingly, the $G$-apparent magnitudes corresponding to the $G$-absolute magnitude of -1 and 4 were determined to be 9 and 14 mag, respectively. Moreover, since the faint limiting magnitude of the photometric data provided by the {\it Gaia} satellite is $G=20.5$ mag \citep[e.g.][]{Gokmen2023, Tasdemir2023, Yontan-Canbay2023}, the faint limiting magnitude calculated in this study can be reliable. Based on the analysis, the number of evolved stars within the completeness limits was determined as 671,600 by applying Equation~\ref{eq: Eq6} to the sample.

\subsection{Space Density Profiles}

To calculate the space density of stars located in a star field direction, space volumes were defined by considering different distance intervals from the Sun, and the number of stars in these volume elements was then determined. The expression used to calculate space density was given in Equation~\ref{eq: Eq7}.
\begin{equation}
D=\frac{N}{\Delta V_{1,2}}
\label{eq: Eq7}
\end{equation}
Here, $D$ stands for the star density, while $N$ is the number of stars. $\Delta V_{1,2}$ represents the space volume that was calculated using the following equation.

\begin{equation}
\Delta V_{1,2}=\left(\frac{\square}{3}\right)\left(\frac{\pi}{180}\right)^2 \left[d_2^3 - d_1^3\right]
\label{eq: Eq8}
\end{equation}
Here, $\square$ is the size of the selected star field (square degree), and $d_1$ and $d_2$ represent two different distances from the Sun, respectively. In this study, to create the density profiles of stars centred on the Sun and located in the space volume $d\leq 1$ kpc, the evolved stars were divided into 200 pc distance intervals from the Sun. As stellar distances increase, space volumes grow, leading to underestimated calculations of space densities. To simultaneously show high and low stellar space densities, the relation $D^{*}=\log D+10$ as presented in the literature, was utilized in this study \citep[c.f.][]{Fenkart1987}. Also, the centroid distance ($d^*$) of the partial volume ($\Delta V_{1,2}$) corresponding to distances $d_1-d_2$ was used to generate the stellar density profiles:
\begin{equation}
d^*=\sqrt[3]{\frac{d_1^3 + d_2^3}{2}}
\label{eq: Eq9}
\end{equation}

\subsection{Galaxy Model Parameters}
\begin{figure}
\centering
\includegraphics[width=\columnwidth]{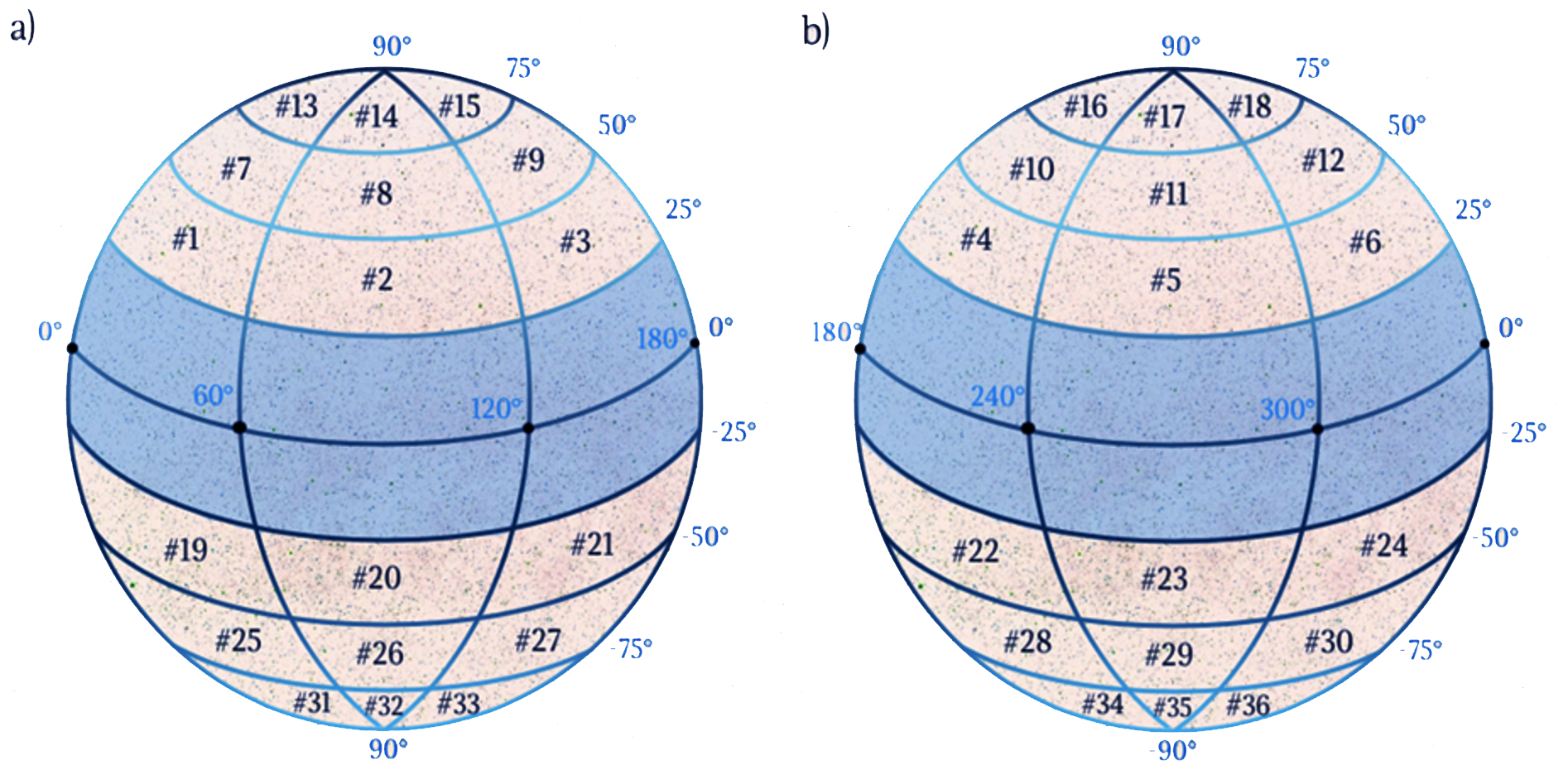}
\caption{Heliocentric celestial sphere was separated into 36-star fields based on the Galactic coordinates of evolved stars: (a)~$0^{\circ}<l\leq180^{\circ}$ and (b) $180^{\circ}< l \leq360^{\circ}$.} 
\label{fig:Fig06}
\end{figure}

This study aims to derive the parameters of the Galaxy model as a function of the Galactic coordinates ($l, b$) and the absolute magnitude ($M_{\rm G}$). To accomplish this, evolved stars were segmented on a heliocentric celestial sphere categorized according to their Galactic latitude and longitude. The selected Galactic latitude intervals were for north and south Galactic hemispheres were $25^{\circ}<|b|\leq50^{\circ}$, $50^{\circ}<|b|\leq75^{\circ}$ and $75^{\circ}<|b|\leq90^{\circ}$, and selected Galactic longitude intervals were $0^{\circ}< l \leq60^{\circ}$, $60^{\circ}< l \leq120^{\circ}$, $120^{\circ}< l \leq180^{\circ}$, $180^{\circ}< l \leq240^{\circ}$, $240^{\circ}< l \leq300^{\circ}$ and $300^{\circ}< l \leq360^{\circ}$. These choices created 18-star fields in the northern Galactic hemisphere and 18-star fields in the southern Galactic hemisphere, totalling the 36-star fields shown in Figure~\ref{fig:Fig06}. The numbers starting with the number sign $\#$ in Figure~\ref{fig:Fig06} give the number of the star field. Since the study focuses on evolved stars located within 1 kpc of the Sun, stars situated in star fields with Galactic latitude $|b|\leq25^\circ$ can reach a maximum distance of approximately 425 pc from the Galactic plane. This calculated value is slightly larger than the one scale height reported for the thin disk in the literature~\citep[c.f.][]{Karaali2004, Bilir2006a} To accurately and precisely determine the scale height of a Galactic population, stars in the star field must be within three to five scale heights of the Galactic plane. For this reason, the part of the sky with $|b|\leq 25^\circ$ and shown in blue on Figure~\ref{fig:Fig06} were removed from the sample, leaving 241,956 evolved stars. 

\begin{figure*}
\centering
\includegraphics[width=2\columnwidth]{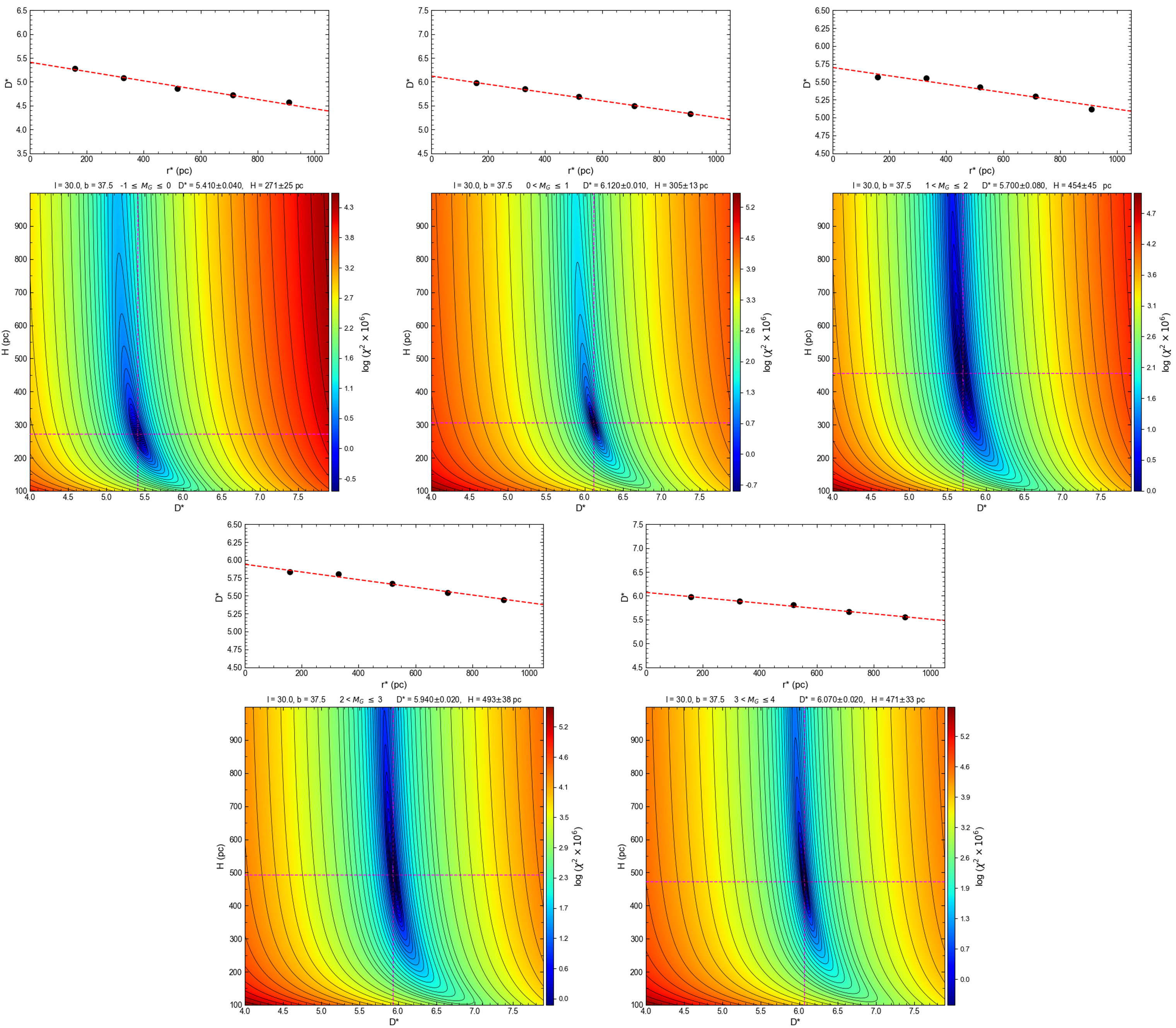}
\caption{Stellar density profiles for five absolute magnitude intervals in star field \#01 (upper panels) and variations of model parameters with $\chi^2$ values (lower panels). The red lines in the upper panels show the Galaxy model fitted to the star density points, and the intersections of the red dashed lines represent the Galaxy model parameters with the minimum $\chi_{\rm min}^2$ value. The numerical values between the two panels were the central coordinates of the star field in the Galactic coordinate system, the absolute magnitude interval, the estimated space density and the scale height.} 
\label{fig:Fig07}
\end{figure*}

To plot the space-density profiles of the star fields, the evolved stars in the sample were divided into five absolute magnitude intervals such $-1<M_{\rm G}~{\rm (mag)}\leq0$, $0<M_{\rm G}~{\rm (mag)}\leq1$, $1<M_{\rm G}~{\rm (mag)}\leq2$, $2<M_{\rm G}~{\rm (mag)}\leq3$, and $3<M_{\rm G}~{\rm (mag)}\leq4$. Evolved stars in each absolute magnitude interval were classified according to their distance, divided into 200 pc distance intervals. The star space density in each distance range was obtained using Equation~\ref{eq: Eq7}. The space volumes were then calculated using Equation~\ref{eq: Eq8} and the sizes of star fields for three different latitude zones as $25^{\circ}<|b|\leq50^{\circ}$, $50^{\circ}<|b|\leq75^{\circ}$ and $75^{\circ}<|b|\leq90^{\circ}$ were calculated as about 1181, 687 and 117 degree$^2$, respectively. In total, 180 space density profiles were constructed within 36-star fields for five consecutive absolute magnitude intervals. The space density profiles are fitted with a single-component density law to obtain the Galaxy model parameters of each star field.

Galaxy model parameters were determined by fitting the density law for the thin-disk population in Equation~\ref{eq: Eq3} to the observational density profiles calculated from the star field. To calculate the Galaxy model parameters, space densities were used in steps of 0.01 in the range $4<D^{*}<8$, and for the scale height, steps of 1 pc in the interval $100<H {\rm (pc)}<1000$. In this study, the parameters of the Galaxy model with the minimum chi-square ($\chi^2_{\rm min}$) were considered in selecting the model that best represents the space-density profile. 

This study requires modelling many star fields and space density profiles obtained from stars of five consecutive absolute magnitude intervals in these fields. To save space in the paper, the analyses were shown in Figure~\ref{fig:Fig07} for evolved stars found at five absolute magnitude intervals in the star field \#01. The Galaxy model parameters estimated for the star fields as a result of the comparison of the observational space densities with the Galaxy model were listed in Table~\ref{tab:parameter_table}. Uncertainties of the estimated space density ($D^*$) and scale height ($H$) for the thin-disk population were given for $\pm 1\sigma$. Moreover, the estimated Galaxy model parameters for the 36-star fields were also listed in Table~\ref{tab:parameter_table}.

\section{Discussion and Conclusion}

\begin{table*}[]
\centering
\caption{Scale heights estimated for five different absolute magnitude intervals in the 36-star fields.}\label{table:Table1}
\resizebox{\textwidth}{!}{%
\begin{tabular}{cccccccccccc}
\hline
& \multicolumn{5}{c}{Absolute Magnitude Intervals}  & &\multicolumn{5}{c}{Absolute Magnitude Intervals}\\ \cline{2-6} \cline{8-12} 
& $-1<M_{\rm G}\leq0$ & $0<M_{\rm G}\leq1$ & $1<M_{\rm G}\leq2$ & $2<M_{\rm G}\leq3$ & $3<M_{\rm G}\leq4$ &                 & $-1<M_{\rm G}\leq0$ & $0<M_{\rm G}\leq1$ & $1<M_{\rm G}\leq2$ & $2<M_{\rm G}\leq3$ & $3<M_{\rm G}\leq4$ \\ \hline
Field           & $H$ (pc)            & $H$ (pc)           & $H$ (pc)           & $H$ (pc)           & $H$ (pc)           & Field           & $H$ (pc)            & $H$ (pc)           & $H$ (pc)           & $H$ (pc)           & $H$ (pc)           \\ \hline
\#01            & 271$\pm$25          & 305$\pm$13         & 454$\pm$45         & 493$\pm$38         & 471$\pm$33         & \#19            & 288$\pm$28          & 351$\pm$19         & 392$\pm$31         & 440$\pm$05         & 555$\pm$38         \\
\#02            & 269$\pm$29          & 287$\pm$15         & 382$\pm$31         & 450$\pm$17         & 411$\pm$26         & \#20            & 254$\pm$08          & 323$\pm$16         & 397$\pm$38         & 487$\pm$05         & 440$\pm$35         \\
\#03            & 242$\pm$08          & 292$\pm$48         & 315$\pm$04         & 371$\pm$24         & 344$\pm$03         & \#21            & 323$\pm$14          & 289$\pm$12         & 340$\pm$26         & 431$\pm$04         & 366$\pm$03         \\
\#04            & 277$\pm$19          & 293$\pm$12         & 271$\pm$17         & 350$\pm$33         & 305$\pm$17         & \#22            & 274$\pm$08          & 301$\pm$16         & 352$\pm$29         & 413$\pm$04         & 315$\pm$09         \\
\#05            & 251$\pm$07          & 313$\pm$02         & 319$\pm$21         & 362$\pm$03         & 341$\pm$20         & \#23            & 295$\pm$10          & 334$\pm$15         & 448$\pm$45         & 396$\pm$27         & 394$\pm$32         \\
\#06            & 264$\pm$07          & 339$\pm$19         & 396$\pm$32         & 405$\pm$01         & 395$\pm$20         & \#24            & 335$\pm$40          & 374$\pm$18         & 401$\pm$05         & 408$\pm$04         & 418$\pm$21         \\
\#07            & 272$\pm$28          & 287$\pm$11         & 368$\pm$09         & 342$\pm$02         & 405$\pm$21         & \#25            & 300$\pm$27          & 307$\pm$05         & 384$\pm$08         & 458$\pm$04         & 476$\pm$03         \\
\#08            & 252$\pm$21          & 298$\pm$13         & 365$\pm$23         & 431$\pm$26         & 382$\pm$21         & \#26            & 225$\pm$06          & 288$\pm$02         & 365$\pm$09         & 394$\pm$10         & 428$\pm$18         \\
\#09            & 238$\pm$24          & 284$\pm$40         & 308$\pm$16         & 382$\pm$35         & 377$\pm$17         & \#27            & 221$\pm$06          & 295$\pm$14         & 357$\pm$35         & 364$\pm$03         & 406$\pm$29         \\
\#10            & 272$\pm$27          & 289$\pm$14         & 324$\pm$22         & 377$\pm$20         & 357$\pm$19         & \#28            & 258$\pm$07          & 259$\pm$02         & 343$\pm$24         & 392$\pm$03         & 357$\pm$02         \\
\#11            & 238$\pm$07          & 283$\pm$11         & 317$\pm$18         & 383$\pm$20         & 393$\pm$24         & \#29            & 299$\pm$08          & 325$\pm$30         & 394$\pm$36         & 462$\pm$40         & 442$\pm$27         \\
\#12            & 253$\pm$01          & 297$\pm$15         & 346$\pm$23         & 389$\pm$34         & 392$\pm$09         & \#30            & 357$\pm$11          & 351$\pm$19         & 404$\pm$09         & 462$\pm$04         & 452$\pm$18         \\
\#13            & 201$\pm$05          & 292$\pm$39         & 315$\pm$22         & 378$\pm$08         & 433$\pm$40         & \#31            & 220$\pm$07          & 280$\pm$13         & 422$\pm$12         & 319$\pm$05         & 392$\pm$08         \\
\#14            & 214$\pm$05          & 295$\pm$47         & 331$\pm$22         & 393$\pm$21         & 495$\pm$45         & \#32            & 253$\pm$10          & 319$\pm$30         & 278$\pm$02         & 435$\pm$38         & 435$\pm$25         \\
\#15            & 220$\pm$07          & 300$\pm$14         & 362$\pm$20         & 459$\pm$29         & 487$\pm$13         & \#33            & 201$\pm$05          & 274$\pm$02         & 288$\pm$06         & 392$\pm$36         & 426$\pm$10         \\
\#16            & 253$\pm$10          & 274$\pm$11         & 348$\pm$35         & 419$\pm$11         & 414$\pm$22         & \#34            & 273$\pm$08          & 276$\pm$12         & 318$\pm$07         & 364$\pm$09         & 404$\pm$35         \\
\#17            & 294$\pm$09          & 281$\pm$41         & 274$\pm$02         & 378$\pm$32         & 363$\pm$18         & \#35            & 246$\pm$06          & 296$\pm$14         & 361$\pm$27         & 353$\pm$19         & 484$\pm$04         \\
\#18            & 219$\pm$22          & 269$\pm$10         & 355$\pm$35         & 379$\pm$09         & 389$\pm$15         & \#36            & 204$\pm$05          & 272$\pm$12         & 299$\pm$05         & 489$\pm$13         & 439$\pm$11         \\ \hline
\textbf{Median} & 253$\pm$10          & 292$\pm$14         & 339$\pm$22         & 383$\pm$21         & 393$\pm$20         & \textbf{Median} & 266$\pm$08          & 299$\pm$14         & 363$\pm$18         & 411$\pm$05         & 427$\pm$18         \\ \hline
\end{tabular}%
}
\end{table*}

In this study, we determined the space densities and scale heights of the Galactic thin-disk population by analyzing the spatial distribution of evolved stars within a heliocentric volume extending to 1 kpc. For this purpose, a sample of 671,600 evolved stars was selected based on their positions in the Hertzsprung-Russell (HR) diagram, using photometric and astrometric data from the {\it Gaia} DR3 catalogue. Each star in the sample had a relative parallax error of less than 0.02, ensuring reliable spatial measurements.

In this study, symmetric star fields from the northern and southern Galactic zones were jointly analyzed to refine the interpretation of the Galaxy model parameters calculated across 36-star fields and within five absolute magnitude intervals. Initially, the space density values derived for the thin disk within symmetric zones were prioritized. To achieve this, the variation in space densities of star fields as a function of their absolute magnitudes in the $D^{*} \times M_{\rm G}$ planes is illustrated on the left side of Figure~\ref{fig:scale_height_all}. Each panel also displays the luminosity functions of evolved stars within a 100-pc space volume \citep{Gaia-Smart2021}. The space densities ($D^{*}$) corresponding to the absolute magnitude intervals of $-1<M_{\rm G}~{\rm (mag)} \leq 0$, $0<M_{\rm G}~{\rm (mag)} \leq 1$, $1<M_{\rm G}~{\rm (mag)} \leq 2$, $2<M_{\rm G}~{\rm (mag)} \leq 3$, and $3<M_{\rm G}~{\rm (mag)} \leq 4$ were calculated as 5.41, 6.09, 5.76, 5.91 and 6.07, respectively, based on the relation $D^{*}=\log D+10$. Overall, the calculated space densities for these five absolute magnitude intervals across star fields within three distinct Galactic latitude zones align closely with the luminosity function obtained for evolved stars by \citet{Gaia-Smart2021}. The consistency between space densities calculated within a 1 kpc space volume and those in existing luminosity function literature further indicates that the estimated scale height for the thin-disk population remains unaffected by parameter degeneracy.

\begin{figure}
\centering
\includegraphics[width=\columnwidth]{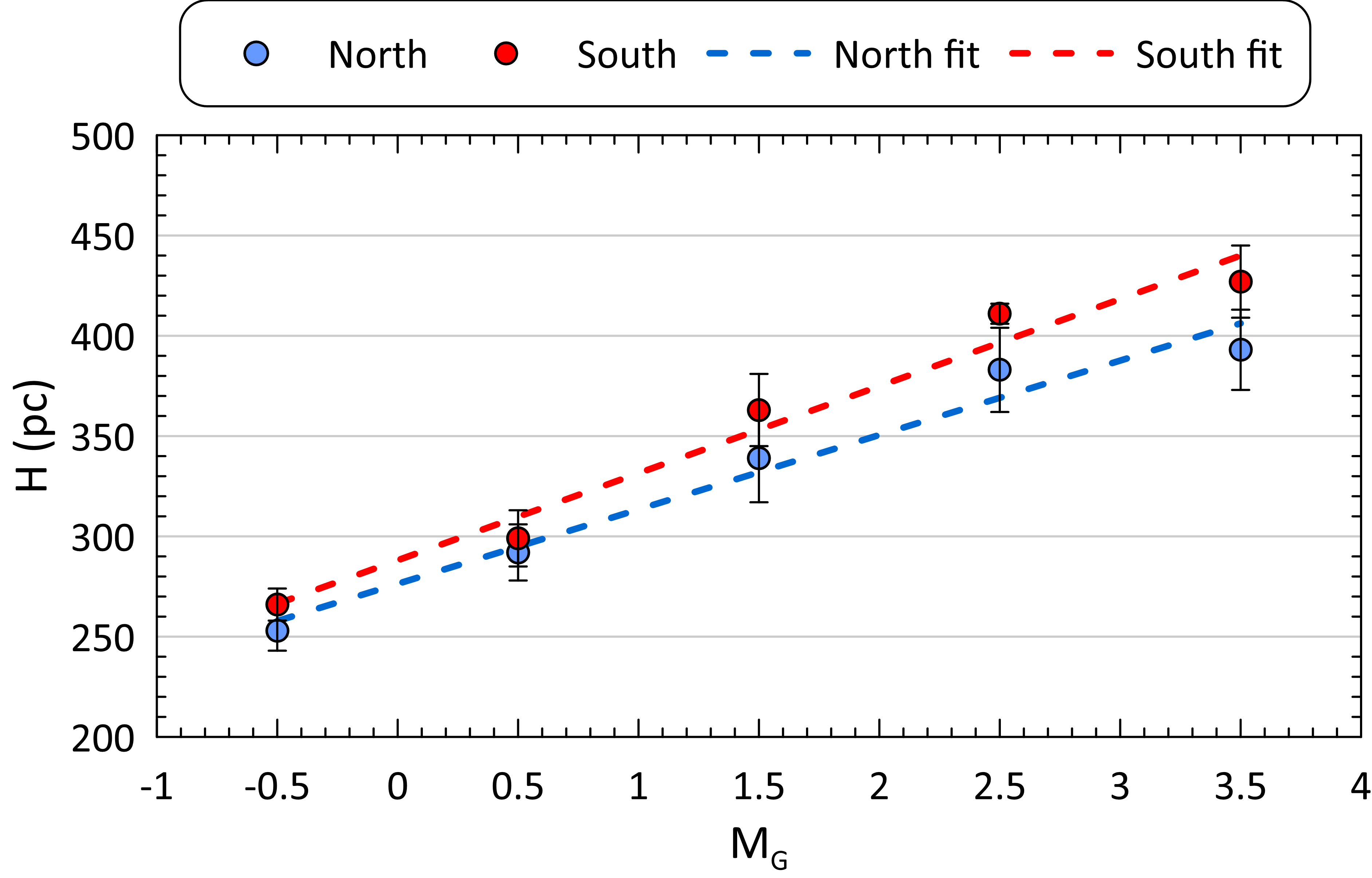}
\caption{The variation in scale heights calculated for evolved stars concerning their absolute magnitudes across both Galactic hemispheres.}
\label{fig:Fig10}
\end{figure}

The Galaxy model parameters of the thin disk were estimated under the assumption that the evolved stars in a 1 kpc space volume predominantly belong to the thin disk in this study. However, it is well-established that the solar neighborhood contains stars from different Galactic populations. To quantify the potential impact of these populations on the results, a Monte Carlo simulation was conducted. For this simulation, a star field was selected for testing and modelled using a single exponential law. This model assumes that the local space density of the thick disk varies $0< n~(\%) \leq15$, and the scale height of the thick disk varies $550\leq H~{\rm (pc)}\leq 1500$. For the halo population, it is assumed that the local space density is between 0.1\% and 0.2\%. These parameter ranges are consistent with values reported in the literature and are listed in Table 1 provided by \citet{Bilir2006a}. The analysis focused on the space densities in the absolute magnitude range $0<M_{\rm G}~{\rm (mag)} \leq 1$, dominated by red-clump stars in the selected star field (\#01). Using a Monte Carlo simulation with 10,000 trials, the Galaxy model parameters for the thick disk and halo populations were randomly selected in each trial. Observational data were subsequently remodelled by reconstructing the exponential density law of the thin disk. The results of the simulation reveal that the scale height of the thin-disk population varies in the range $295\leq H~({\rm pc}) \leq 310$, with the most probable value determined as $H=303\pm 3$ pc based on a Gaussian fit to the distribution. This value is in excellent agreement with the $305 \pm 13$ pc obtained in this study without discriminating between different stellar populations (see Table~\ref{table:Table1}). These findings indicate that the scale height of the thin disk, as derived from evolved stars within a 1 kpc space volume, is minimally affected by the presence of stars from other Galactic populations. This result underscores the robustness of the adopted method and provides a reliable estimate for the thin disk’s vertical structure.

The variation in scale heights of evolved stars in the thin-disk population is analysed to their Galactic coordinates and absolute magnitude intervals. The methodological steps used to interpret space density parameters have likewise been applied to the determination of the scale height parameter. For this purpose, the scale heights estimated for evolved stars in the five absolute magnitude intervals covering star fields in both the northern and southern Galactic hemispheres were provided in Table~\ref{table:Table1}. The right side of Figure~\ref{fig:scale_height_all} illustrates the range of scale heights for different absolute magnitude intervals in these symmetric star fields, which spans from $200<H~{\rm(pc)}< 600$. Notably, an increasing trend is observed in scale heights from brighter to fainter absolute magnitudes within each Galactic latitude zone, a trend that similarly appears for symmetric star fields at equivalent latitudes. This trend is further evident from the median scale height values calculated for each absolute magnitude interval in the bottom row of Table~\ref{table:Table1}. The variation in median scale heights calculated for evolved stars concerning their absolute magnitudes in both Galactic hemispheres is presented in Figure \ref{fig:Fig10}. It has been seen that, as the absolute magnitudes of stars transition from brighter to fainter, the scale height increases from 250 to 430 pc. In addition, the scale heights calculated for evolved stars in the southern hemisphere are slightly larger than those obtained for stars in the northern hemisphere. Considering the trend observed in the data from both hemispheres, linear fits have been applied, resulting in the following relations. 

\begin{eqnarray}
    H_{\rm North} = 37.1\times M_{\rm G} + 276~~~~~~(R^2=0.968) \nonumber \\ 
    H_{\rm South} = 43.4\times M_{\rm G} + 288~~~~~~(R^2=0.970)
\end{eqnarray}
The relationships calculated for both Galactic hemispheres exhibit a high degree of correlation. Although the slopes of the two relationships are nearly identical, there is a difference in the zero point of 12 pc between the southern and northern Galactic hemispheres.

It can be seen from Figure~\ref{fig:Fig02} that the red clump stars, which stand out as a dense population among the evolved stars analyzed in this study, are located in the absolute magnitude $0<M_{\rm G}~{\rm (mag)} \leq 1$ interval. The scale heights of red clump stars in the northern and southern Galactic hemispheres were also estimated as $H_{\rm north} = 292\pm 14$ and $H_{\rm south} = 299\pm 14$ pc, respectively, as given in Table~\ref{table:Table1}. When the scale heights of the red clump stars in both Galactic hemispheres and the weighted average of their errors were calculated, the scale height of the red clump stars in the solar neighbourhood was determined to be $H=295\pm 10$ pc. Although this result is slightly larger than the scale height values of $150<H~{\rm (pc)}<300$ obtained by \citet{Cabrera-Lavers2007}, who estimated the Galaxy model parameters for the thin-disk population from 2MASS photometric data of red clump stars at high Galactic latitudes, it is in good agreement with the scale height of $H=280$ pc determined by \citet{Bovy2016b} from their analysis of 14,699 red clump stars selected from the APOGEE survey.

\begin{figure}
\centering
\includegraphics[width=\columnwidth]{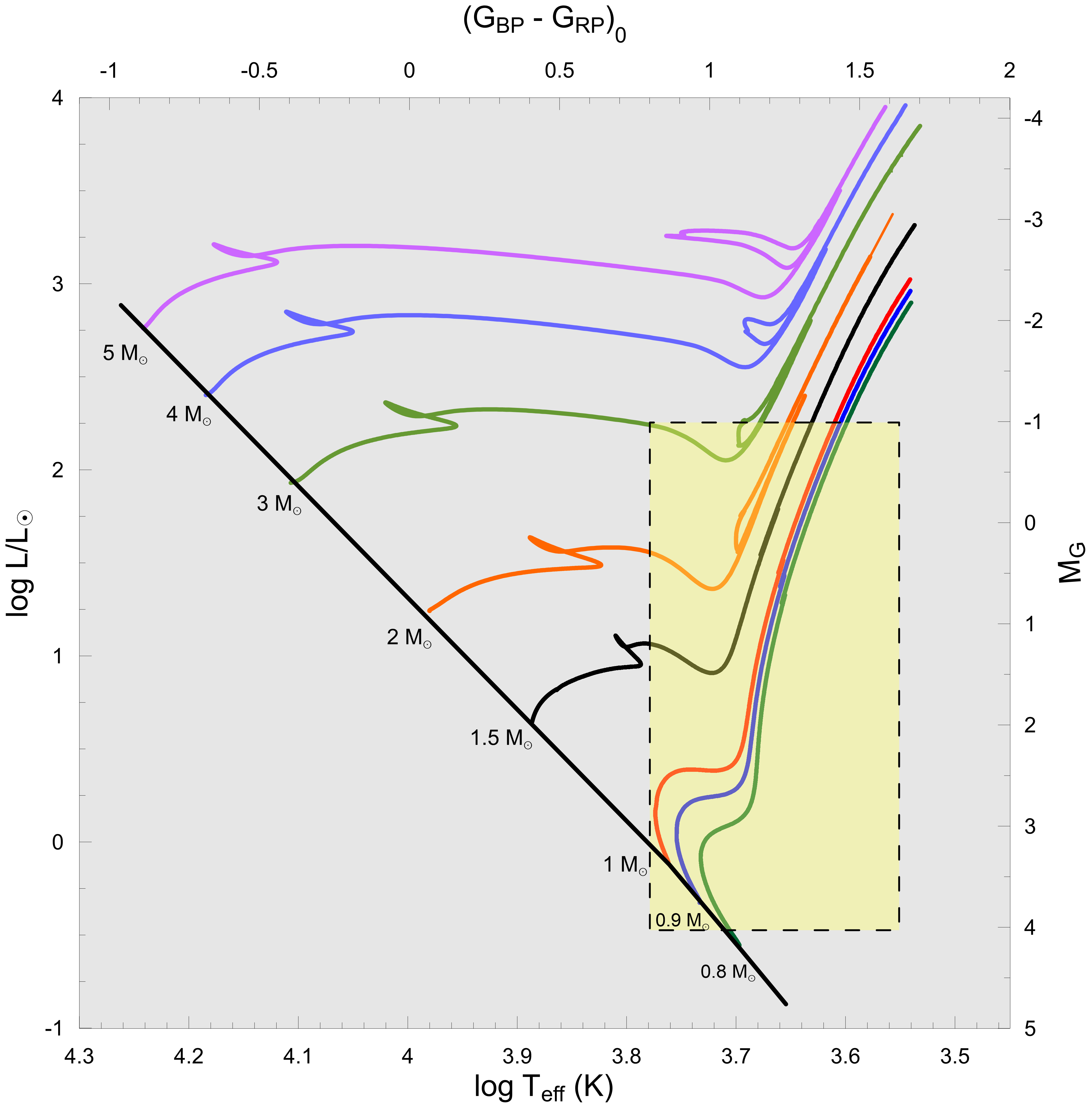}
\caption{The positions of the PARSEC mass tracks of main-sequence stars of different masses on the CMD. The curves with different colors indicate different masses, and dashed lines indicate the region where the evolved stars studied were located.}
\label{fig:Parsec-HR}
\end{figure}

The space densities obtained from these star fields align closely with those observed in the solar neighbourhood. However, the observed dependency of scale height on Galactic coordinates and absolute magnitude intervals indicates the need for further investigation to clarify the underlying causes of this variation. In this study, the observed variations in scale height were not attributed to parameter degeneracy, as demonstrated by the consistency of space densities with those of the solar neighbourhood. The smaller scale heights of evolved stars with bright absolute magnitudes indicate that these systems predominantly consist of younger stars. To address this analysis, we employed the Padova and Trieste Stellar Evolution Code (PARSEC) stellar evolution models \citep{Bressan2012, Tang2014, Chen2015}. Specifically, we considered mass tracks for solar-metallicity stars with heavy element abundance $Z=0.014$ and helium abundance $Y=0.273$. From these stellar evolution models, we selected eight theoretical stars with masses ranging from $0.8\leq M/M_{\odot}\leq5$. These theoretical stars, which evolved from the main sequence to advanced evolutionary stages, were depicted on the $M_{\rm G} \times (G_{\rm BP}-G_{\rm RP})_0$ CDM as shown in Figure~\ref{fig:Parsec-HR}. The associated luminosity ($L/L_{\odot}$) and temperature ($\log T_{\rm eff}$) were displayed on the upper and right axes of the CMD. Based on the {\it Gaia} color index and the absolute magnitudes of the evolved stars analyzed in this study, and their alignment with the PARSEC mass tracks, we concluded that these stars reside within the mass range $0.8<M/M_{\odot}<3$, corresponding to spectral types K3 and B8.5, respectively \citep[c.f.][]{Eker2015, Eker2018, Eker2020, Eker2024}.

In the literature, the relations between the spectral types of stars and their scale height have been studied by many researchers \citep[c.f.][]{Gilmore-Reid1983, Hawkins1988, Pirzkal2005, Kong2008, Holwerda2014}. These studies have shown that thin-disk scale heights increase from 100 pc to 400 pc as it moves from O spectral type main-sequence stars to M spectral type ones. Considering that the scale heights of the evolved stars analyzed in this study range between 200 and 600 pc (see Figure~\ref{fig:scale_height_all}), it is seen that the values given for stars of spectral type F and K were in agreement. This finding, together with the duration of their stay in the main sequence and the scale heights in the literature, explains the reason for the small-scale heights of bright absolute magnitude stars and the large-scale heights of faint absolute magnitude stars, as well.

In this study, we utilized the high-precision photometric and astrometric data from the {\it Gaia} satellite to investigate the stellar number density and scale heights of the thin-disk population in the solar neighbourhood across various Galactic coordinates and absolute magnitude intervals. Our analysis reveals that the thin-disk space densities align well with previously reported values for a 100 pc space volume. In contrast, the scale heights of evolved stars across different star fields vary substantially, ranging from 200 to 600 pc, increasing from bright absolute magnitudes ($-1<M_{\rm G}~{\rm (mag)}\leq 0$) to faint ones ($3<M_{\rm G}~{\rm (mag)}\leq 4$). This large variation in scale heights for evolved thin-disk stars over a limited range of absolute magnitudes likely reflects the evolutionary effects of stars with different masses in the solar neighbourhood. Specifically, the alignment of scale heights for bright absolute magnitudes with those of early-type main-sequence stars, supported by mass track data, underscores that evolved stars exhibit varied scale heights based on their evolutionary stages. Additionally, our study confirms that Galaxy model parameters, when derived from {\it Gaia} data, vary with Galactic coordinates, object luminosities, and space volumes \citep[e.g.][]{Bilir2006a, Bilir2006b, Bilir2006c, Cabrera-Lavers2007, Ak2007a, Yaz2010, Yaz2015, Karaali2007}. The results obtained in this study using evolved stars suggest that the Galaxy model parameters should be evaluated in terms of Galactic coordinates and the absolute magnitudes of the stars.

\section*{Acknowledgments}
We sincerely thank the referee for their insightful and constructive suggestions, which have significantly contributed to improving the quality and clarity of the manuscript. The authors express their gratitude to Prof. Dr. Salih Karaali for their valuable insights and inspiration that contributed to the development of this paper. We would like to thank Hikmet Çakmak for his contribution to the codes used in the analyses. This research has made use of NASA's Astrophysics Data System Bibliographic Services.
\emph{Gaia}\footnote{https://www.cosmos.esa.int/gaia}, processed by the \emph{Gaia} Data Processing and Analysis Consortium (DPAC)\footnote{https://www.cosmos.esa.int/web/gaia/dpac/consortium}. Funding for DPAC has been provided by national institutions, in particular, the institutions participating in the \emph{Gaia} Multilateral Agreement.

\software{python \citep{python3}, astropy \citep{astropy:2013, astropy:2018, astropy:2022}, numpy \citep{numpy:2020}, matplotlib \citep{matplotlib:2007}, mwdust \citep{Bovy2016a}
          }

\bibliography{Galactic_Structure}{}
\bibliographystyle{aasjournal}

\appendix
\restartappendixnumbering

\section{Additional Figures and Tables}
\begin{longtable}{ccccc|ccccc}
\caption{Galaxy model parameters estimated for five different absolute magnitude intervals in the 36-star fields.}
\label{tab:parameter_table}\\
\hline
\multicolumn{5}{c|}{$0^{\circ}< l \leq60^{\circ}$, $25^{\circ}< b \leq50^{\circ}$, Field $\#$01}      & \multicolumn{5}{c}{$60^{\circ}< l \leq120^{\circ}$, $25^{\circ}< b \leq50^{\circ}$, Field $\#$02}    \\ \hline
\endfirsthead
\multicolumn{10}{c}%
{{\bfseries Table \thetable\ continued from previous page}} \\
\endhead
\hline
\endfoot
\endlastfoot
$M_1-M_2$       & $N$        & $D^*$              & $H$ (pc)       & $\chi_{\rm min}^2 (10^{-7})$     & $M_1-M_2$          & $N$        & $D^*$              & $H$ (pc)        & $\chi_{\rm min}^2 (10^{-7})$      \\ \hline
(-1, 0{]}       & 628        & 5.41$\pm$0.04      & 271$\pm$25     & 1.79                             & (-1, 0{]}          & 563        & 5.38$\pm$0.04      & 269$\pm$29      & 3.05                              \\
(0, 1{]}        & 3764       & 6.12$\pm$0.01      & 305$\pm$13     & 1.70                             & (0, 1{]}           & 3539       & 6.13$\pm$0.02      & 287$\pm$15      & 3.30                              \\
(1, 2{]}        & 2208       & 5.70$\pm$0.08      & 454$\pm$45     & 10.07                            & (1, 2{]}           & 1996       & 5.73$\pm$0.02      & 382$\pm$31      & 7.75                              \\
(2, 3{]}        & 4179       & 5.94$\pm$0.02      & 493$\pm$38     & 7.75                             & (2, 3{]}           & 3612       & 5.91$\pm$0.01      & 450$\pm$17      & 0.33                              \\
(3, 4{]}        & 5453       & 6.07$\pm$0.02      & 471$\pm$33     & 2.96                             & (3, 4{]}           & 4468       & 6.04$\pm$0.02      & 411$\pm$26      & 2.47                              \\ \hline
\multicolumn{5}{c|}{$120^{\circ}< l \leq180^{\circ}$, $25^{\circ}< b \leq50^{\circ}$, Field $\#$03}   & \multicolumn{5}{c}{$180^{\circ}< l \leq240^{\circ}$, $25^{\circ}< b \leq50^{\circ}$, Field $\#$04}   \\ \hline
$M_1-M_2$       & $N$        & $D^*$              & $H$ (pc)       & $\chi_{\rm min}^2 (10^{-7})$     & $M_1-M_2$          & $N$        & $D^*$              & $H$ (pc)        & $\chi_{\rm min}^2 (10^{-7})$      \\ \hline
(-1, 0{]}       & 521        & 5.41$\pm$0.01      & 242$\pm$8      & 0.93                             & (-1, 0{]}          & 489        & 5.42$\pm$0.04      & 277$\pm$19      & 6.63                              \\
(0, 1{]}        & 3078       & 6.07$\pm$0.06      & 292$\pm$48     & 29.00                            & (0, 1{]}           & 3014       & 6.05$\pm$0.01      & 293$\pm$12      & 2.53                              \\
(1, 2{]}        & 1690       & 5.75$\pm$0.01      & 315$\pm$4      & 10.37                            & (1, 2{]}           & 1445       & 5.77$\pm$0.03      & 271$\pm$17      & 2.50                              \\
(2, 3{]}        & 3162       & 5.94$\pm$0.02      & 371$\pm$24     & 2.49                             & (2, 3{]}           & 2958       & 5.94$\pm$0.06      & 350$\pm$33      & 11.70                             \\
(3, 4{]}        & 3666       & 6.04$\pm$0.01      & 344$\pm$3      & 0.60                             & (3, 4{]}           & 3180       & 6.04$\pm$0.02      & 305$\pm$17      & 8.21                              \\ \hline
\multicolumn{5}{c|}{$240^{\circ}< l \leq300^{\circ}$, $25^{\circ}< b \leq50^{\circ}$, Field $\#$05}   & \multicolumn{5}{c}{$300^{\circ}< l \leq360^{\circ}$, $25^{\circ}< b \leq50^{\circ}$, Field $\#$06}   \\ \hline
$M_1-M_2$       & $N$        & $D^*$              & $H$ (pc)       & $\chi_{\rm min}^2 (10^{-7})$     & $M_1-M_2$          & $N$        & $D^*$              & $H$ (pc)        & $\chi_{\rm min}^2 (10^{-7})$      \\ \hline
(-1, 0{]}       & 500        & 5.37$\pm$0.01      & 251$\pm$7      & 0.41                             & (-1, 0{]}          & 578        & 5.45$\pm$0.01      & 264$\pm$7       & 0.82                              \\
(0, 1{]}        & 3793       & 6.11$\pm$0.01      & 313$\pm$2      & 0.81                             & (0, 1{]}           & 3715       & 6.06$\pm$0.02      & 339$\pm$19      & 4.11                              \\
(1, 2{]}        & 1803       & 5.78$\pm$0.02      & 319$\pm$21     & 3.18                             & (1, 2{]}           & 2213       & 5.76$\pm$0.02      & 396$\pm$32      & 8.05                              \\
(2, 3{]}        & 3352       & 5.98$\pm$0.01      & 362$\pm$3      & 0.42                             & (2, 3{]}           & 4015       & 6.00$\pm$0.01      & 405$\pm$1       & 0.09                              \\
(3, 4{]}        & 3641       & 6.04$\pm$0.02      & 341$\pm$20     & 1.31                             & (3, 4{]}           & 4939       & 6.10$\pm$0.01      & 395$\pm$20      & 1.59                              \\ \hline
\multicolumn{5}{c|}{$0^{\circ}< l \leq60^{\circ}$, $50^{\circ}< b \leq75^{\circ}$, Field $\#$07}      & \multicolumn{5}{c}{$60^{\circ}< l \leq120^{\circ}$, $50^{\circ}< b \leq75^{\circ}$, Field $\#$08}    \\ \hline
$M_1-M_2$       & $N$        & $D^*$              & $H$ (pc)       & $\chi_{\rm min}^2 (10^{-7})$     & $M_1-M_2$          & $N$        & $D^*$              & $H$ (pc)        & $\chi_{\rm min}^2 (10^{-7})$      \\ \hline
(-1, 0{]}       & 184        & 5.40$\pm$0.05      & 272$\pm$28     & 3.48                             & (-1,0{]}           & 187        & 5.46$\pm$0.04      & 252$\pm$21      & 7.96                              \\
(0, 1{]}        & 1124       & 6.13$\pm$0.02      & 287$\pm$11     & 6.44                             & (0, 1{]}           & 1170       & 6.12$\pm$0.02      & 298$\pm$13      & 4.29                              \\
(1, 2{]}        & 726        & 5.76$\pm$0.01      & 368$\pm$9      & 0.72                             & (1, 2{]}           & 691        & 5.75$\pm$0.03      & 365$\pm$23      & 9.71                              \\
(2, 3{]}        & 1357       & 6.08$\pm$0.01      & 342$\pm$2      & 0.61                             & (2, 3{]}           & 1231       & 5.89$\pm$0.02      & 431$\pm$26      & 2.27                              \\
(3, 4{]}        & 1905       & 6.11$\pm$0.02      & 405$\pm$21     & 3.29                             & (3, 4{]}           & 1715       & 6.10$\pm$0.02      & 382$\pm$21      & 6.17                              \\ \hline
\multicolumn{5}{c|}{$120^{\circ}< l \leq180^{\circ}$, $50^{\circ}< b \leq75^{\circ}$, Field $\#$09}   & \multicolumn{5}{c}{$180^{\circ}< l \leq240^{\circ}$, $50^{\circ}< b \leq75^{\circ}$, Field $\#$10}   \\ \hline
$M_1-M_2$       & $N$        & $D^*$              & $H$ (pc)       & $\chi_{\rm min}^2 (10^{-7})$     & $M_1-M_2$          & $N$        & $D^*$              & $H$ (pc)        & $\chi_{\rm min}^2 (10^{-7})$      \\ \hline
(-1, 0{]}       & 184        & 5.41$\pm$0.13      & 238$\pm$24     & 15.40                            & (-1, 0{]}          & 173        & 5.36$\pm$0.05      & 272$\pm$27      & 1.07                              \\
(0, 1{]}        & 1059       & 6.12$\pm$0.06      & 284$\pm$40     & 10.60                            & (0, 1{]}           & 1033       & 6.09$\pm$0.02      & 289$\pm$14      & 6.63                              \\
(1, 2{]}        & 609        & 5.81$\pm$0.03      & 308$\pm$16     & 6.96                             & (1, 2{]}           & 578        & 5.75$\pm$0.03      & 324$\pm$22      & 2.32                              \\
(2, 3{]}        & 1183       & 5.95$\pm$0.07      & 382$\pm$35     & 11.50                            & (2, 3{]}           & 1243       & 5.98$\pm$0.02      & 377$\pm$20      & 9.16                              \\
(3, 4{]}        & 1445       & 6.04$\pm$0.02      & 377$\pm$17     & 2.21                             & (3, 4{]}           & 1307       & 6.03$\pm$0.02      & 357$\pm$19      & 9.50                              \\ \hline
\multicolumn{5}{c|}{$240^{\circ}< l \leq300^{\circ}$, $50^{\circ}< b \leq75^{\circ}$, Field $\#$11}   & \multicolumn{5}{c}{$300^{\circ}< l \leq360^{\circ}$, $50^{\circ}< b \leq75^{\circ}$, Field $\#$12}   \\ \hline
$M_1-M_2$       & $N$        & $D^*$              & $H$ (pc)       & $\chi_{\rm min}^2 (10^{-7})$     & $M_1-M_2$          & $N$        & $D^*$              & $H$ (pc)        & $\chi_{\rm min}^2 (10^{-7})$      \\ \hline
(-1, 0{]}       & 144        & 5.41$\pm$0.02      & 238$\pm$7      & 0.29                             & (-1, 0{]}          & 179        & 5.45$\pm$0.01      & 253$\pm$1       & 0.02                              \\
(0, 1{]}        & 1055       & 6.12$\pm$0.02      & 283$\pm$11     & 1.39                             & (0, 1{]}           & 1021       & 6.07$\pm$0.02      & 297$\pm$15      & 1.62                              \\
(1, 2{]}        & 602        & 5.79$\pm$0.03      & 317$\pm$18     & 1.09                             & (1, 2{]}           & 646        & 5.77$\pm$0.03      & 346$\pm$23      & 8.09                              \\
(2, 3{]}        & 1112       & 5.92$\pm$0.02      & 383$\pm$20     & 1.08                             & (2, 3{]}           & 1235       & 5.96$\pm$0.07      & 389$\pm$34      & 15.30                             \\
(3, 4{]}        & 1505       & 6.03$\pm$0.02      & 393$\pm$24     & 4.20                             & (3, 4{]}           & 1748       & 6.10$\pm$0.01      & 392$\pm$9       & 0.97                              \\ \hline
\multicolumn{5}{c|}{$0^{\circ}< l \leq60^{\circ}$, $75^{\circ}< b \leq90^{\circ}$, Field $\#$13}      & \multicolumn{5}{c}{$60^{\circ}< l \leq120^{\circ}$, $75^{\circ}< b \leq90^{\circ}$, Field $\#$14}    \\ \hline
$M_1-M_2$       & $N$        & $D^*$              & $H$ (pc)       & $\chi_{\rm min}^2 (10^{-7})$     & $M_1-M_2$          & $N$        & $D^*$              & $H$ (pc)        & $\chi_{\rm min}^2 (10^{-7})$      \\ \hline
(-1, 0{]}       & 22         & 5.66$\pm$0.02      & 201$\pm$5      & 0.17                             & (-1, 0{]}          & 25         & 5.64$\pm$0.02      & 214$\pm$5       & 0.35                              \\
(0, 1{]}        & 157        & 6.13$\pm$0.06      & 292$\pm$39     & 38.00                            & (0, 1{]}           & 139        & 6.06$\pm$0.07      & 295$\pm$47      & 12.90                             \\
(1, 2{]}        & 70         & 5.72$\pm$0.03      & 315$\pm$22     & 1.44                             & (1, 2{]}           & 79         & 5.72$\pm$0.03      & 331$\pm$22      & 1.85                              \\
(2, 3{]}        & 150        & 5.90$\pm$0.01      & 378$\pm$8      & 0.51                             & (2, 3{]}           & 159        & 5.90$\pm$0.02      & 393$\pm$21      & 1.01                              \\
(3, 4{]}        & 252        & 6.03$\pm$0.06      & 433$\pm$40     & 84.9                             & (3, 4{]}           & 282        & 6.02$\pm$0.07      & 495$\pm$45      & 71.90                             \\ \hline
\multicolumn{5}{c|}{$120^{\circ}< l \leq180^{\circ}$, $75^{\circ}< b \leq90^{\circ}$, Field $\#$15}   & \multicolumn{5}{c}{$180^{\circ}< l \leq240^{\circ}$, $75^{\circ}< b \leq90^{\circ}$, Field $\#$16}   \\ \hline
$M_1-M_2$       & $N$        & $D^*$              & $H$ (pc)       & $\chi_{\rm min}^2 (10^{-7})$     & $M_1-M_2$          & $N$        & $D^*$              & $H$ (pc)        & $\chi_{\rm min}^2 (10^{-7})$      \\ \hline
(-1, 0{]}       & 11         & 5.33$\pm$0.02      & 220$\pm$7      & 0.35                             & (-1, 0{]}          & 17         & 5.31$\pm$0.02      & 253$\pm$10      & 0.63                              \\
(0, 1{]}        & 150        & 6.08$\pm$0.02      & 300$\pm$14     & 8.28                             & (0, 1{]}           & 132        & 6.11$\pm$0.02      & 274$\pm$11      & 1.22                              \\
(1, 2{]}        & 109        & 5.79$\pm$0.03      & 362$\pm$20     & 5.20                             & (1, 2{]}           & 98         & 5.80$\pm$0.09      & 348$\pm$35      & 49.60                             \\
(2, 3{]}        & 197        & 5.90$\pm$0.02      & 459$\pm$29     & 9.76                             & (2, 3{]}           & 180        & 5.88$\pm$0.01      & 419$\pm$11      & 0.13                              \\
(3, 4{]}        & 240        & 5.94$\pm$0.01      & 487$\pm$13     & 0.18                             & (3, 4{]}           & 213        & 5.99$\pm$0.02      & 414$\pm$22      & 8.95                              \\ \hline
\multicolumn{5}{c|}{$240^{\circ}< l \leq300^{\circ}$, $75^{\circ}< b \leq90^{\circ}$, Field $\#$17}   & \multicolumn{5}{c}{$300^{\circ}< l \leq360^{\circ}$, $75^{\circ}< b \leq90^{\circ}$, Field $\#$18}   \\ \hline
$M_1-M_2$       & $N$        & $D^*$              & $H$ (pc)       & $\chi_{\rm min}^2 (10^{-7})$     & $M_1-M_2$          & $N$        & $D^*$              & $H$ (pc)        & $\chi_{\rm min}^2 (10^{-7})$      \\ \hline
(-1, 0{]}       & 22         & 5.26$\pm$0.01      & 294$\pm$9      & 0.40                             & (-1, 0{]}          & 13         & 5.39$\pm$0.20      & 219$\pm$22      & 17.40                             \\
(0, 1{]}        & 132        & 6.09$\pm$0.07      & 281$\pm$41     & 31.20                            & (0, 1{]}           & 137        & 6.12$\pm$0.02      & 269$\pm$10      & 3.65                              \\
(1, 2{]}        & 75         & 5.87$\pm$0.01      & 274$\pm$2      & 0.10                             & (1, 2{]}           & 89         & 5.74$\pm$0.10      & 355$\pm$35      & 17.30                             \\
(2, 3{]}        & 160        & 5.94$\pm$0.08      & 378$\pm$32     & 17.3                             & (2, 3{]}           & 151        & 5.90$\pm$0.01      & 379$\pm$9       & 0.77                              \\
(3, 4{]}        & 205        & 6.07$\pm$0.02      & 363$\pm$18     & 5.10                             & (3, 4{]}           & 247        & 6.10$\pm$0.02      & 389$\pm$15      & 3.30                              \\ \hline
\multicolumn{5}{c|}{$0^{\circ}< l \leq60^{\circ}$, $-50^{\circ}\leq b <-25^{\circ}$, Field $\#$19}    & \multicolumn{5}{c}{$60^{\circ}< l \leq120^{\circ}$, $-50^{\circ}\leq b <-25^{\circ}$, Field $\#$20}  \\ \hline
$M_1-M_2$       & $N$        & $D^*$              & $H$ (pc)       & $\chi_{\rm min}^2 (10^{-7})$     & $M_1-M_2$          & $N$        & $D^*$              & $H$ (pc)        & $\chi_{\rm min}^2 (10^{-7})$      \\ \hline
(-1, 0{]}       & 662        & 5.40$\pm$0.04      & 288$\pm$28     & 2.24                             & (-1, 0{]}          & 596        & 5.43$\pm$0.01      & 254$\pm$8       & 0.76                              \\
(0, 1{]}        & 4558       & 6.13$\pm$0.02      & 351$\pm$19     & 2.61                             & (0, 1{]}           & 4035       & 6.12$\pm$0.02      & 323$\pm$16      & 8.72                              \\
(1, 2{]}        & 2175       & 5.76$\pm$0.02      & 392$\pm$31     & 9.69                             & (1, 2{]}           & 2148       & 5.75$\pm$0.08      & 397$\pm$38      & 11.90                             \\
(2, 3{]}        & 3390       & 5.89$\pm$0.01      & 440$\pm$5      & 0.29                             & (2, 3{]}           & 3730       & 5.89$\pm$0.01      & 487$\pm$5       & 0.56                              \\
(3, 4{]}        & 6402       & 6.07$\pm$0.01      & 555$\pm$38     & 1.92                             & (3, 4{]}           & 5212       & 6.07$\pm$0.05      & 440$\pm$35      & 13.30                             \\ \hline
\multicolumn{5}{c|}{$120^{\circ}< l \leq180^{\circ}$, $-50^{\circ}\leq b <-25^{\circ}$, Field $\#$21} & \multicolumn{5}{c}{$180^{\circ}< l \leq240^{\circ}$, $-50^{\circ}\leq b <-25^{\circ}$, Field $\#$22} \\ \hline
$M_1-M_2$       & $N$        & $D^*$              & $H$ (pc)       & $\chi_{\rm min}^2 (10^{-7})$     & $M_1-M_2$          & $N$        & $D^*$              & $H$ (pc)        & $\chi_{\rm min}^2 (10^{-7})$      \\ \hline
(-1, 0{]}       & 495        & 5.31$\pm$0.01      & 323$\pm$14     & 0.83                             & (-1, 0{]}          & 610        & 5.40$\pm$0.01      & 274$\pm$8       & 0.11                              \\
(0, 1{]}        & 3265       & 6.09$\pm$0.02      & 289$\pm$12     & 1.95                             & (0, 1{]}           & 3438       & 6.09$\pm$0.02      & 301$\pm$16      & 1.33                              \\
(1, 2{]}        & 1839       & 5.76$\pm$0.02      & 340$\pm$26     & 9.77                             & (1, 2{]}           & 1958       & 5.76$\pm$0.03      & 352$\pm$29      & 2.57                              \\
(2, 3{]}        & 3481       & 5.91$\pm$0.01      & 431$\pm$4      & 0.13                             & (2, 3{]}           & 3336       & 5.91$\pm$0.01      & 413$\pm$4       & 0.25                              \\
(3, 4{]}        & 4230       & 6.07$\pm$0.00      & 366$\pm$3      & 0.39                             & (3, 4{]}           & 3531       & 6.07$\pm$0.01      & 315$\pm$9       & 0.58                              \\ \hline
\multicolumn{5}{c|}{$240^{\circ}< l \leq300^{\circ}$, $-50^{\circ}\leq b <-25^{\circ}$, Field $\#$23} & \multicolumn{5}{c}{$300^{\circ}< l \leq360^{\circ}$, $-50^{\circ}\leq b <-25^{\circ}$, Field $\#$24} \\ \hline
$M_1-M_2$       & $N$        & $D^*$              & $H$ (pc)       & $\chi_{\rm min}^2 (10^{-7})$     & $M_1-M_2$          & $N$        & $D^*$              & $H$ (pc)        & $\chi_{\rm min}^2 (10^{-7})$      \\ \hline
(-1, 0{]}       & 706        & 5.41$\pm$0.01      & 295$\pm$10     & 0.41                             & (-1, 0{]}          & 749        & 5.37$\pm$0.04      & 335$\pm$40      & 1.67                              \\
(0, 1{]}        & 3974       & 6.09$\pm$0.01      & 334$\pm$15     & 0.81                             & (0, 1{]}           & 4829       & 6.12$\pm$0.01      & 374$\pm$18      & 1.64                              \\
(1, 2{]}        & 2366       & 5.73$\pm$0.02      & 448$\pm$45     & 3.18                             & (1, 2{]}           & 2395       & 5.78$\pm$0.01      & 401$\pm$5       & 0.43                              \\
(2, 3{]}        & 2997       & 5.88$\pm$0.02      & 396$\pm$27     & 0.42                             & (2, 3{]}           & 3687       & 5.91$\pm$0.01      & 408$\pm$4       & 0.67                              \\
(3, 4{]}        & 4695       & 6.07$\pm$0.05      & 394$\pm$32     & 1.31                             & (3, 4{]}           & 6534       & 6.09$\pm$0.01      & 418$\pm$21      & 1.52                              \\ \hline
\multicolumn{5}{c|}{$0^{\circ}< l \leq60^{\circ}$,$-75^{\circ}\leq b <-50^{\circ}$, Field $\#$25}     & \multicolumn{5}{c}{$60^{\circ}< l \leq120^{\circ}$, $-75^{\circ}\leq b <-50^{\circ}$, Field $\#$26}  \\ \hline
$M_1-M_2$       & $N$        & $D^*$              & $H$ (pc)       & $\chi_{\rm min}^2 (10^{-7})$     & $M_1-M_2$          & $N$        & $D^*$              & $H$ (pc)        & $\chi_{\rm min}^2 (10^{-7})$      \\ \hline
(-1, 0{]}       & 223        & 5.40$\pm$0.05      & 300$\pm$27     & 1.30                             & (-1, 0{]}          & 130        & 5.42$\pm$0.02      & 225$\pm$6       & 1.67                              \\
(0, 1{]}        & 1144       & 6.09$\pm$0.01      & 307$\pm$5      & 0.76                             & (0, 1{]}           & 1034       & 6.10$\pm$0.01      & 288$\pm$2       & 0.66                              \\
(1, 2{]}        & 766        & 5.75$\pm$0.01      & 384$\pm$8      & 0.36                             & (1, 2{]}           & 700        & 5.75$\pm$0.01      & 365$\pm$9       & 0.31                              \\
(2, 3{]}        & 1411       & 5.91$\pm$0.01      & 458$\pm$4      & 0.66                             & (2, 3{]}           & 1154       & 5.91$\pm$0.01      & 394$\pm$10      & 0.16                              \\
(3, 4{]}        & 2155       & 6.07$\pm$0.02      & 476$\pm$3      & 0.95                             & (3, 4{]}           & 1954       & 6.09$\pm$0.01      & 428$\pm$18      & 1.39                              \\ \hline
\multicolumn{5}{c|}{$120^{\circ}< l \leq180^{\circ}$, $-75^{\circ}\leq b <-50^{\circ}$, Field $\#$27} & \multicolumn{5}{c}{$180^{\circ}< l \leq240^{\circ}$, $-75^{\circ}\leq b <-50^{\circ}$, Field $\#$28} \\ \hline
$M_1-M_2$       & $N$        & $D^*$              & $H$ (pc)       & $\chi_{\rm min}^2 (10^{-7})$     & $M_1-M_2$          & $N$        & $D^*$              & $H$ (pc)        & $\chi_{\rm min}^2 (10^{-7})$      \\ \hline
(-1, 0{]}       & 125        & 5.42$\pm$0.02      & 221$\pm$6      & 0.69                             & (-1, 0{]}          & 159        & 5.39$\pm$0.01      & 258$\pm$7       & 0.13                              \\
(0, 1{]}        & 1083       & 6.10$\pm$0.02      & 295$\pm$14     & 1.78                             & (0, 1{]}           & 867        & 6.11$\pm$0.01      & 259$\pm$2       & 0.32                              \\
(1, 2{]}        & 649        & 5.77$\pm$0.01      & 357$\pm$35     & 30.50                            & (1, 2{]}           & 698        & 5.80$\pm$0.03      & 343$\pm$24      & 6.08                              \\
(2, 3{]}        & 1083       & 5.94$\pm$0.01      & 364$\pm$3      & 0.51                             & (2, 3{]}           & 1243       & 5.95$\pm$0.01      & 392$\pm$3       & 0.63                              \\
(3, 4{]}        & 1611       & 6.05$\pm$0.06      & 406$\pm$29     & 15.10                            & (3, 4{]}           & 1517       & 6.10$\pm$0.01      & 357$\pm$2       & 0.89                              \\ \hline
\multicolumn{5}{c|}{$240^{\circ}< l \leq300^{\circ}$, $-75^{\circ}\leq b <-50^{\circ}$, Field $\#$29} & \multicolumn{5}{c}{$300^{\circ}< l \leq360^{\circ}$,$-75^{\circ}\leq b <-50^{\circ}$, Field $\#$30}  \\ \hline
$M_1-M_2$       & $N$        & $D^*$              & $H$ (pc)       & $\chi_{\rm min}^2 (10^{-7})$     & $M_1-M_2$          & $N$        & $D^*$              & $H$ (pc)        & $\chi_{\rm min}^2 (10^{-7})$      \\ \hline
(-1, 0{]}       & 222        & 5.40$\pm$0.01      & 299$\pm$8      & 0.52                             & (-1, 0{]}          & 286        & 5.37$\pm$0.01      & 357$\pm$11      & 0.02                              \\
(0, 1{]}        & 1249       & 6.10$\pm$0.07      & 325$\pm$30     & 1.47                             & (0, 1{]}           & 1300       & 6.05$\pm$0.02      & 351$\pm$19      & 1.62                              \\
(1, 2{]}        & 792        & 5.77$\pm$0.09      & 394$\pm$36     & 1.37                             & (1, 2{]}           & 812        & 5.74$\pm$0.01      & 404$\pm$9       & 8.09                              \\
(2, 3{]}        & 1449       & 5.93$\pm$0.07      & 462$\pm$40     & 14.90                            & (2, 3{]}           & 1362       & 5.89$\pm$0.01      & 462$\pm$4       & 15.30                             \\
(3, 4{]}        & 2008       & 6.07$\pm$0.02      & 442$\pm$27     & 1.03                             & (3, 4{]}           & 2270       & 6.12$\pm$0.01      & 452$\pm$18      & 0.97                              \\ \hline
\multicolumn{5}{c|}{$0^{\circ}< l \leq60^{\circ}$, $-90^{\circ}\leq b <-75^{\circ}$, Field $\#$31}    & \multicolumn{5}{c}{$60^{\circ}< l \leq120^{\circ}$, $-90^{\circ}\leq b <-75^{\circ}$, Field $\#$32}  \\ \hline
$M_1-M_2$       & $N$        & $D^*$              & $H$ (pc)       & $\chi_{\rm min}^2 (10^{-7})$     & $M_1-M_2$          & $N$        & $D^*$              & $H$ (pc)        & $\chi_{\rm min}^2 (10^{-7})$      \\ \hline
(-1, 0{]}       & 13         & 5.33$\pm$0.02      & 220$\pm$7      & 0.35                             & (-1, 0{]}          & 17         & 5.31$\pm$0.02      & 253$\pm$10      & 0.63                              \\
(0, 1{]}        & 141        & 6.13$\pm$0.02      & 280$\pm$13     & 4.99                             & (0, 1{]}           & 155        & 6.07$\pm$0.07      & 319$\pm$30      & 63.40                             \\
(1, 2{]}        & 111        & 5.66$\pm$0.01      & 422$\pm$12     & 0.17                             & (1, 2{]}           & 58         & 5.75$\pm$0.01      & 278$\pm$2       & 0.40                              \\
(2, 3{]}        & 118        & 5.93$\pm$0.01      & 319$\pm$5      & 0.42                             & (2, 3{]}           & 196        & 5.95$\pm$0.08      & 435$\pm$38      & 53.30                             \\
(3, 4{]}        & 263        & 6.10$\pm$0.01      & 392$\pm$8      & 0.66                             & (3, 4{]}           & 262        & 6.06$\pm$0.06      & 435$\pm$25      & 21.00                             \\ \hline
\multicolumn{5}{c|}{$120^{\circ}< l \leq180^{\circ}$, $-90^{\circ}\leq b <-75^{\circ}$, Field $\#$33} & \multicolumn{5}{c}{$180^{\circ}< l \leq240^{\circ}$, $-90^{\circ}\leq b <-75^{\circ}$, Field $\#$34} \\ \hline
$M_1-M_2$       & $N$        & $D^*$              & $H$ (pc)       & $\chi_{\rm min}^2 (10^{-7})$     & $M_1-M_2$          & $N$        & $D^*$              & $H$ (pc)        & $\chi_{\rm min}^2 (10^{-7})$      \\ \hline
(-1, 0{]}       & 22         & 5.66$\pm$0.02      & 201$\pm$5      & 0.17                             & (-1, 0{]}          & 21         & 5.28$\pm$0.02      & 273$\pm$8       & 0.19                              \\
(0, 1{]}        & 133        & 6.12$\pm$0.01      & 274$\pm$2      & 0.75                             & (0, 1{]}           & 134        & 6.11$\pm$0.02      & 276$\pm$12      & 1.68                              \\
(1, 2{]}        & 78         & 5.74$\pm$0.01      & 288$\pm$6      & 0.25                             & (1, 2{]}           & 71         & 5.71$\pm$0.01      & 318$\pm$7       & 0.20                              \\
(2, 3{]}        & 172        & 5.94$\pm$0.07      & 392$\pm$36     & 70.00                            & (2, 3{]}           & 144        & 5.91$\pm$0.01      & 364$\pm$9       & 0.17                              \\
(3, 4{]}        & 239        & 6.02$\pm$0.01      & 426$\pm$10     & 59.20                            & (3, 4{]}           & 230        & 6.05$\pm$0.06      & 404$\pm$35      & 13.30                             \\ \hline
\multicolumn{5}{c|}{$240^{\circ}< l \leq300^{\circ}$, $-90^{\circ}\leq b <-75^{\circ}$, Field $\#$35} & \multicolumn{5}{c}{$300^{\circ}< l \leq360^{\circ}$, $-90^{\circ}\leq b <-75^{\circ}$, Field $\#$36} \\ \hline
$M_1-M_2$       & $N$        & $D^*$              & $H$ (pc)       & $\chi_{\rm min}^2 (10^{-7})$     & $M_1-M_2$          & $N$        & $D^*$              & $H$ (pc)        & $\chi_{\rm min}^2 (10^{-7})$      \\ \hline
(-1, 0{]}       & 32         & 5.60$\pm$0.01      & 246$\pm$6      & 0.84                             & (-1, 0{]}          & 23         & 5.66$\pm$0.01      & 204$\pm$5       & 0.17                              \\
(0, 1{]}        & 150        & 6.10$\pm$0.02      & 296$\pm$14     & 1.40                             & (0, 1{]}           & 133        & 6.13$\pm$0.02      & 272$\pm$12      & 1.95                              \\
(1, 2{]}        & 88         & 5.71$\pm$0.03      & 361$\pm$27     & 2.47                             & (1, 2{]}           & 75         & 5.73$\pm$0.01      & 299$\pm$5       & 0.23                              \\
(2, 3{]}        & 138        & 5.92$\pm$0.02      & 353$\pm$19     & 1.16                             & (2, 3{]}           & 226        & 5.86$\pm$0.01      & 489$\pm$13      & 0.20                              \\
(3, 4{]}        & 290        & 6.01$\pm$0.01      & 484$\pm$4      & 0.24                             & (3, 4{]}           & 288        & 6.08$\pm$0.01      & 439$\pm$11      & 0.72                              \\ \hline
\end{longtable}

\newpage
\pagebreak

\begin{figure}
\gridline{\fig{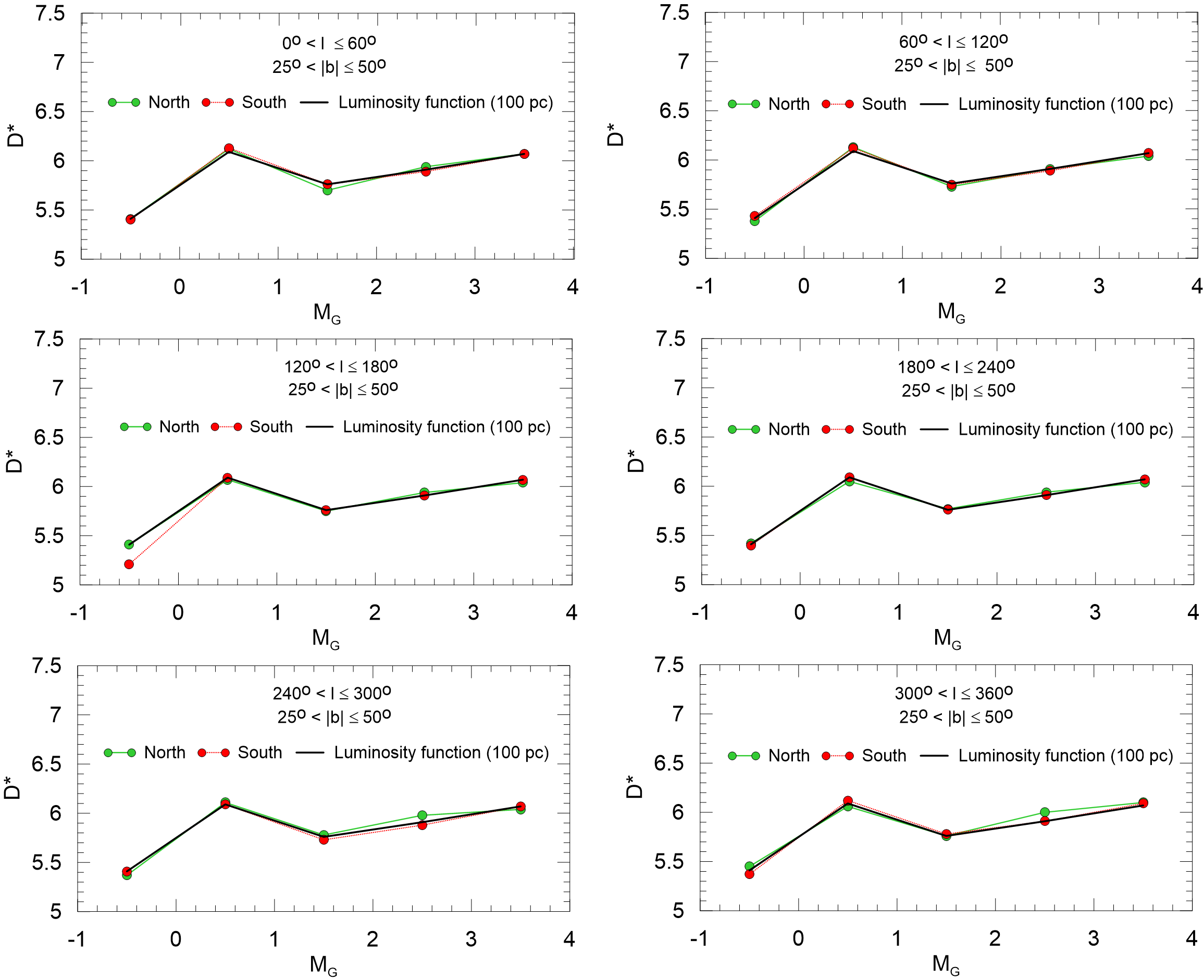}{0.45\textwidth}{}
          \fig{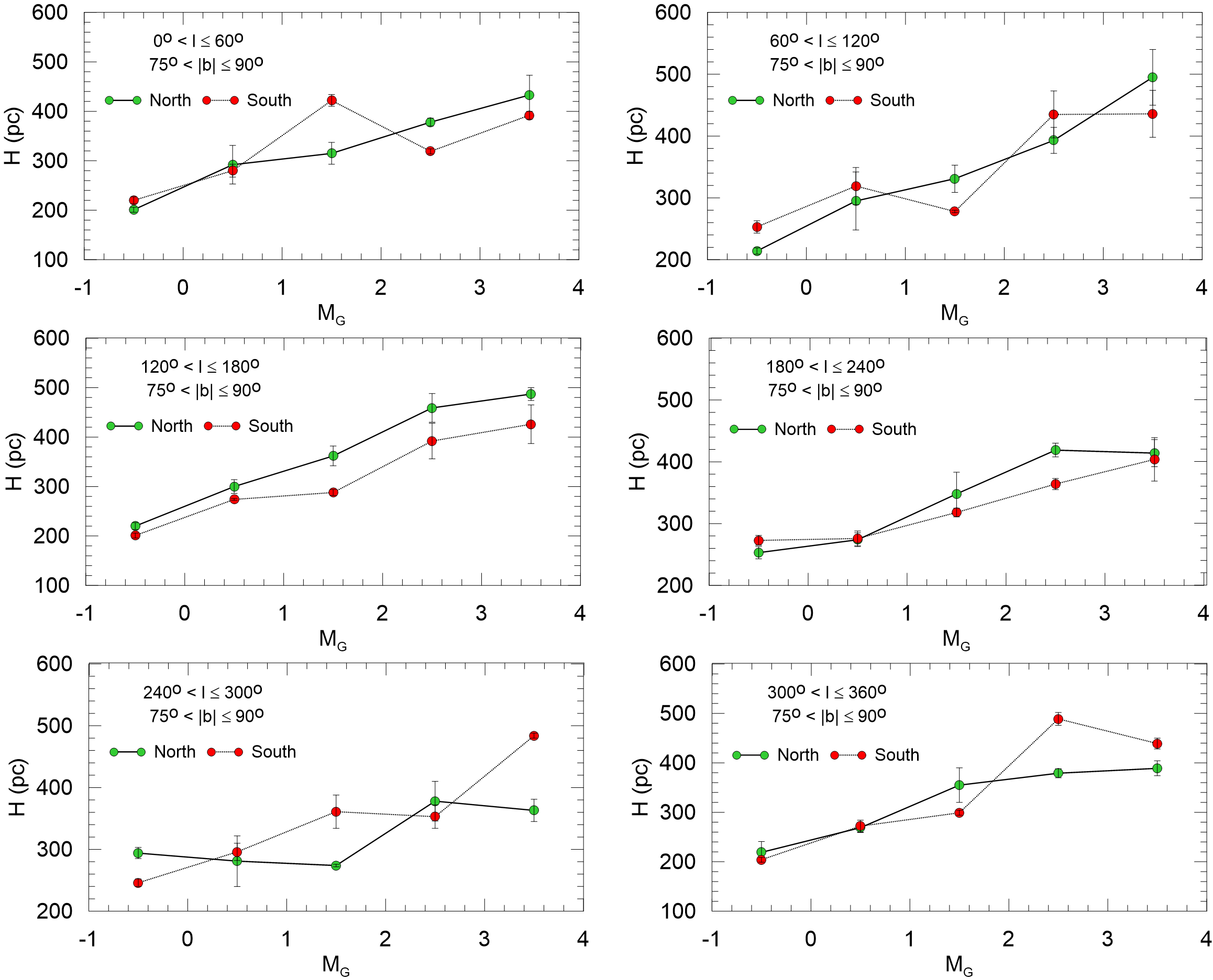}{0.45\textwidth}{}}
\gridline{\fig{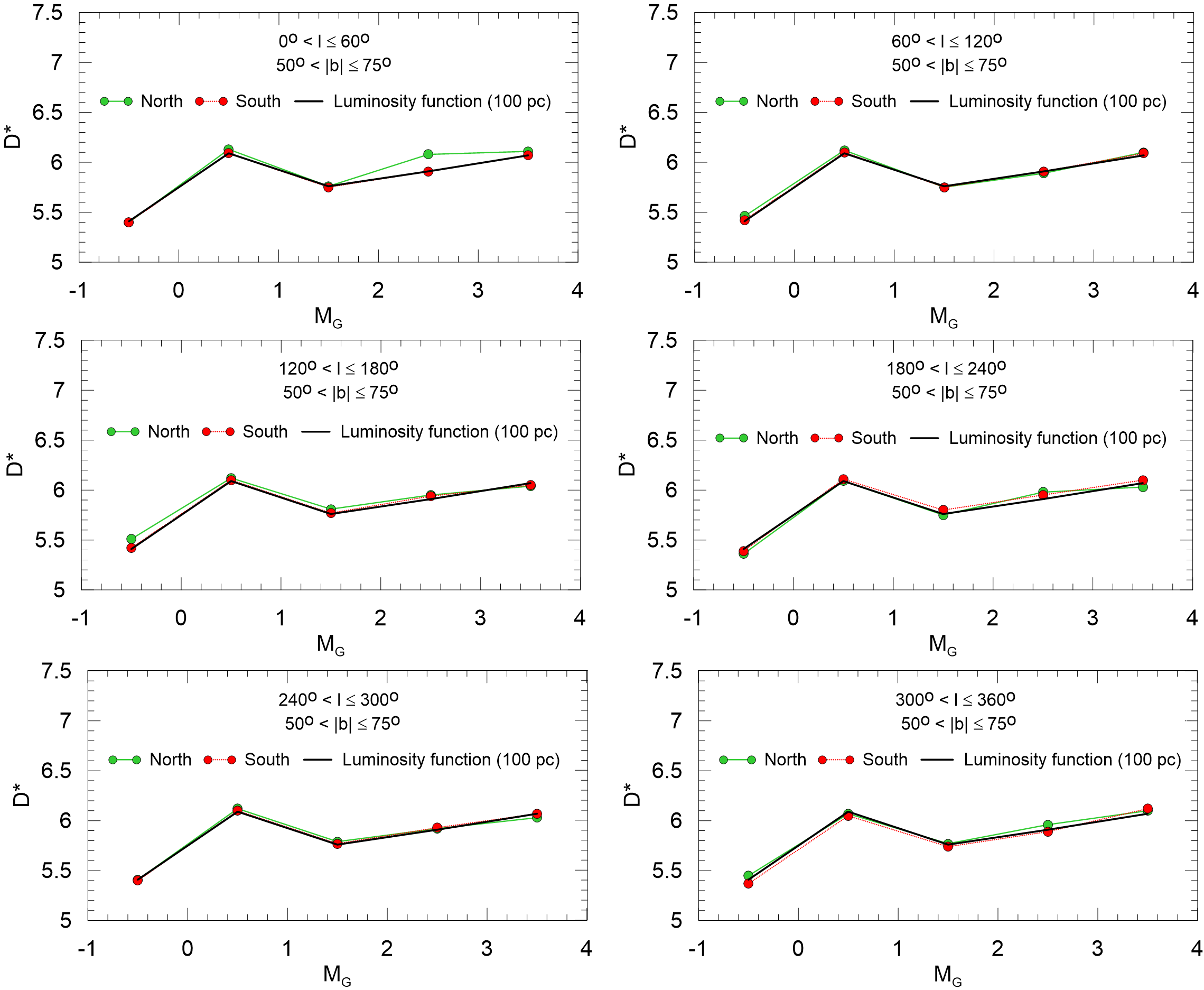}{0.45\textwidth}{}
          \fig{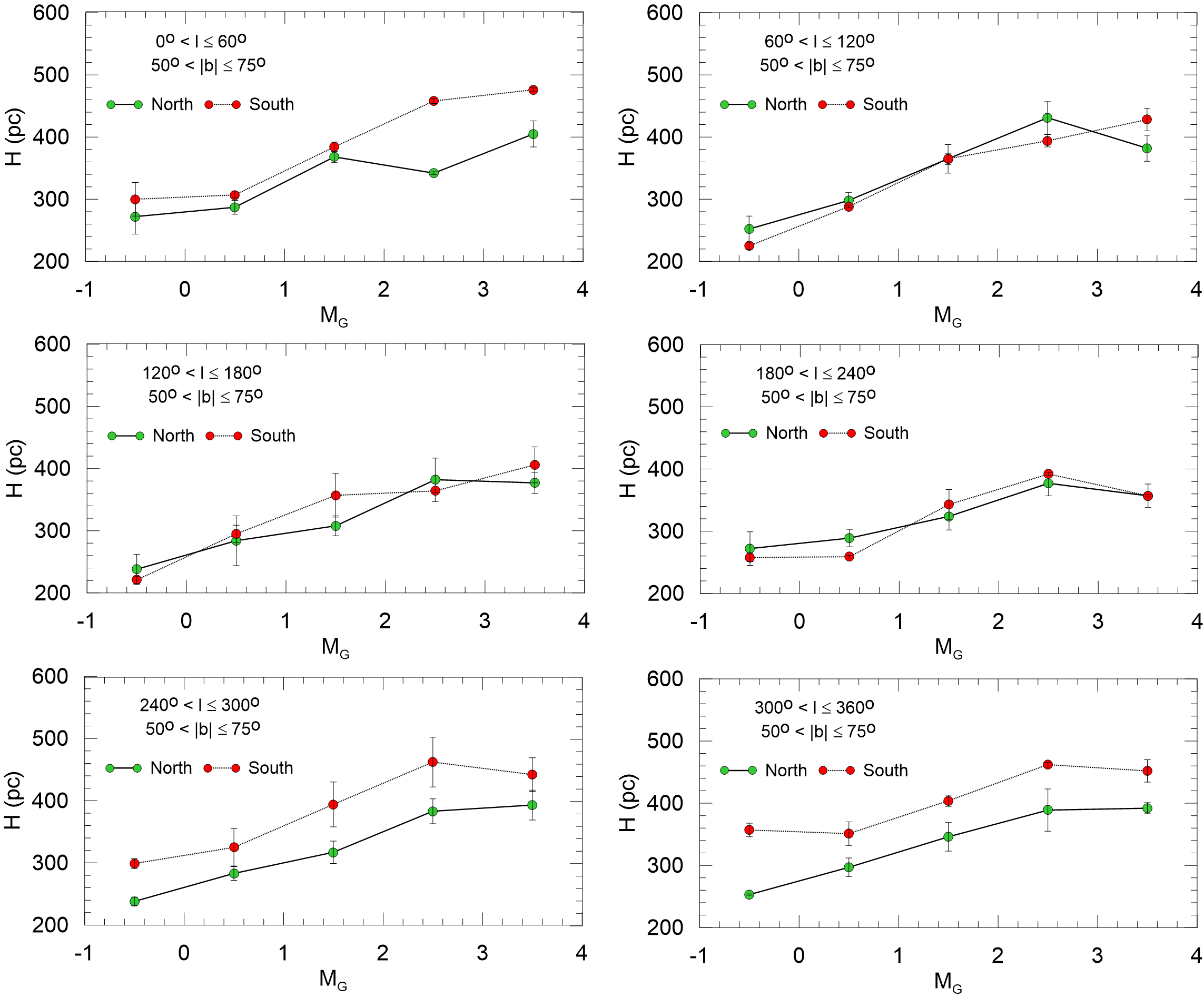}{0.45\textwidth}{}}
\gridline{\fig{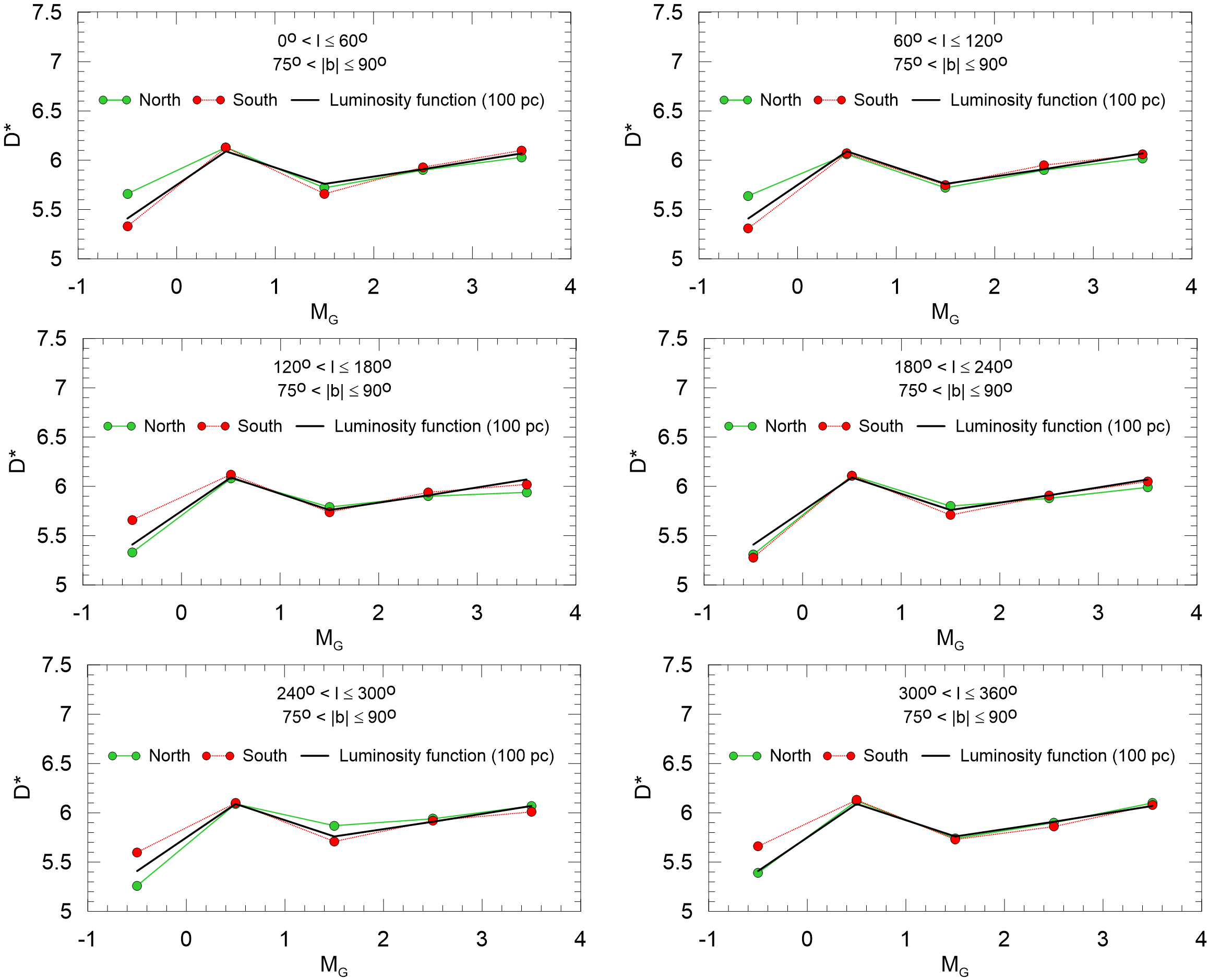}{0.45\textwidth}{}
          \fig{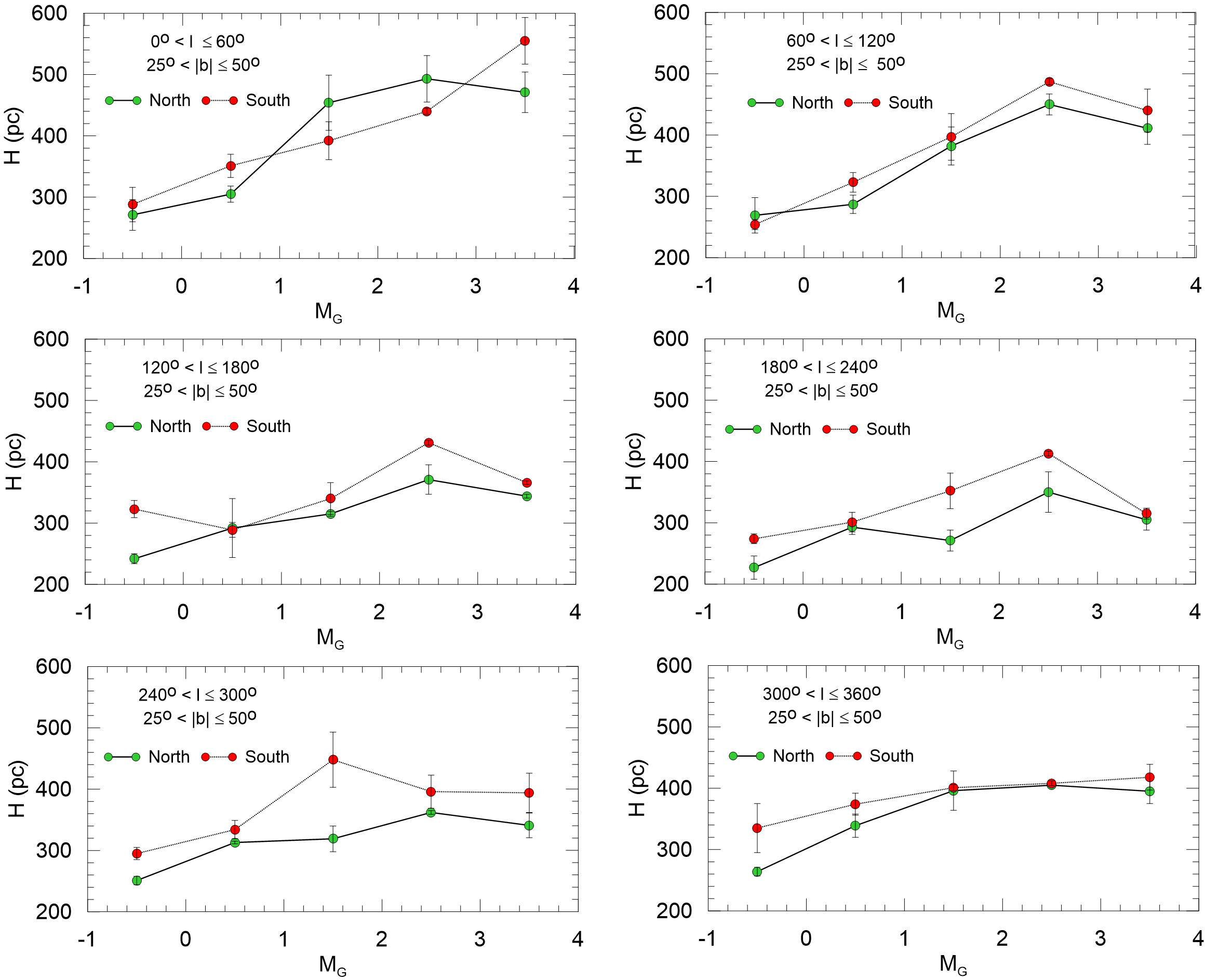}{0.45\textwidth}{}}
\caption{Space densities (left panels) and scale heights (right panels) estimated for different five absolute magnitude intervals of star fields within the Galactic latitude intervals in the 36-star fields. Green and red solid lines represent star fields in the north and south Galactic hemispheres, respectively.}\label{fig:scale_height_all}
\end{figure}







\end{document}